\documentclass[twocolumn,amsmath,aps,fleqn]{revtex4}
 
\usepackage{amssymb}
\usepackage{amsmath}
\usepackage{epsfig}
\usepackage{subfigure}
\usepackage{mathrsfs}
\usepackage{longtable}
\usepackage{enumerate} 
\usepackage{multirow}
\usepackage{color}

\newcommand{\be}{\begin{eqnarray}}
\newcommand{\ee}{\end{eqnarray}}
\newcommand{\Hu}{{\cal H}}

 \allowdisplaybreaks[1]

\begin{document}

\title{Exploring degeneracies in modified gravity with weak lensing}
\author{C. Danielle Leonard}
\email{danielle.leonard@astro.ox.ac.uk}
\affiliation{Astrophysics, University of Oxford, Denys Wilkinson Building, Keble Road, Oxford, OX1 3RH, UK}

\author{Tessa Baker}
\email{tessa.baker@astro.ox.ac.uk}
\affiliation{Astrophysics, University of Oxford, Denys Wilkinson Building, Keble Road, Oxford, OX1 3RH, UK}

\author{Pedro G. Ferreira}
\email{p.ferreira1@physics.ox.ac.uk}
\affiliation{Astrophysics, University of Oxford, Denys Wilkinson Building, Keble Road, Oxford, OX1 3RH, UK}

% ----------------------- ABSTRACT -------------------------
 
\begin{abstract}
By considering linear-order departures from general relativity, we compute a novel expression for the weak lensing convergence power spectrum under alternative theories of gravity. This comprises an integral over a `kernel' of general relativistic quantities multiplied by a theory-dependent `source' term. The clear separation between theory-independent and -dependent terms allows for an explicit understanding of each physical effect introduced by altering the theory of gravity.  We take advantage of this to explore the degeneracies between gravitational parameters in weak lensing observations. 
\end{abstract}

\maketitle

\section{Introduction}
\label{section:intro}
\noindent 

In recent years, weak gravitational lensing has been put forth as a promising method of testing gravitation on cosmological scales
\cite{Schmidt2008, Thomas2009, Amendola2008, Tsujikawa2008, Zhang2007, Bertschinger2008, Knox2006, Ishak2006}, with some exciting first constraints having been found already \cite{Simpson2012,Reyes2010}. Moreover, advances in relevant data analysis (for example, \cite{Miller2012}) and the coming next generation of lensing-optimised surveys mean that we will soon be in a position to take full advantage of the potential of weak lensing.

Stronger constraints on gravity are obtained by combining weak gravitational lensing with other probes. One observable which is commonly touted as providing particularly complementary constraints to weak lensing is $f\sigma_8(a)$.  Here, $f(a)$ is the linear growth rate of structure, defined as:
\begin{equation}
f(a)=\frac{d \ln \Delta_M(a)}{d \ln a}
\label{fofa}
\end{equation}
where $\Delta_M(a)$ is the amplitude of the growing mode of the matter density perturbation, and $\sigma_8(a)$ is the amplitude of the matter power spectrum within spheres of radius $8$ Mpc/h. The combination $f\sigma_8(a)$ can be constrained through measurements of redshift-space distortions in galaxy surveys. 

In \cite{Baker2014}, an expression for $f\sigma_8(a)$ was derived in the case of linear deviations from the model of  general relativity (GR) with $\Lambda$CDM. Here we build on this work by constructing a similar expression for $P_\kappa(\ell)$, the angular power spectrum of the weak lensing observable convergence ($\kappa$). The main advantage of our expression is that it clearly distinguishes the physical source of all modified gravity effects to $P_\kappa(\ell)$, which allows for a more thorough interpretation and understanding of these effects than previously possible. While we focus on $P_\kappa(\ell)$ in this work, recall that the two main weak lensing observables, convergence and shear, can be trivially interconverted \cite{Bartelmann2001}. Therefore, we treat convergence as a proxy for weak lensing more generally, and all expressions which we derive could be equivalently and easily formulated in terms of shear. 

This paper is structured as follows: in Section \ref{section:derivation}, we detail the derivation of the
expression for $P_\kappa(\ell)$.  Section \ref{section:under_deg} discusses how weak lensing degeneracy directions between gravitational parameters can be
understood with the help of our expression. Finally, Section \ref{section:constraints} provides forecast constraints on deviations from GR+$\Lambda$CDM from future
surveys, and interprets these constraints using our expression for $P_\kappa(\ell)$.  We conclude in Section \ref{section:discussion}.

\section{Convergence in modified gravity: the linear response approach}
\label{section:derivation}
\noindent
In what follows, we use the scalar perturbed Friedmann-Robertson-Walker (FRW) metric in the conformal Newtonian gauge, with the following form:
\begin{equation}
ds^2=a(\tau)^2 \left[-(1+2\Psi)d\tau^2+(1-2\Phi)dx_i dx^i\right].
\label{frw}
\end{equation}
Our parameterisation of alternative theories of gravity makes use of the quasistatic approximation (see, for example, \cite{Silvestri2013}). The quasistatic approximation states that within the range of scales relevant for current galaxy surveys, the most significant effects of a sizeable class of modified theories can be captured by introducing two functions of time and scale into the linearised field equations of GR. These functions play the role of a modified gravitational constant, and a non-unity (late-time) ratio of the two scalar gravitational potentials:
\begin{align}
2 \nabla^2 \Phi(a,k) &= 8 \pi G a^2 \mu(a,k) \bar{\rho}_M\Delta_M(a,k)\nonumber \\
\frac{\Phi(a,k)}{\Psi(a,k)}&=\gamma(a,k). 
\label{defgammu}
\end{align}
In GR, both $\gamma(a,k)$ and $\mu(a,k)$ are equal to $1$.  

Clearly equation \ref{defgammu} can only be an effective description of more complicated, exact sets of field equations \cite{Baker:2012zs,Creminelli:2008wc,Baker:2011jy, Battye:2012eu,Gleyzes2013,Gleyzes2014,Bloomfield:2012ff,Amendola_Fogli2013,Hojjati2013,Hojjati2014}. However, several works have numerically verified the validity of the quasistatic approximation in many gravity theories (notably those with one new degree of freedom) on the distance scales considered here \cite{Noller2014, Schmidt2009, Zhao2011, Barreira2013, Li2013}.

We first compute the power spectrum of the convergence in general relativity, and then generalise to alternative theories of gravity. We make the simplifying assumption that radiation can be neglected for all redshifts of interest in this paper. That is, we take $\Omega_M^{GR}(z)+\Omega_\Lambda^{GR}(z)=1$.

\subsection{Calculating convergence: general relativity}
\label{subsection:convergenceGR}

The convergence, $\kappa$, describes the magnification of an image due to lensing. This effect is captured by the geodesic equation for the displacement of a photon transverse to the line of sight. In the cosmological weak lensing context of general relativity, this is given by:
\begin{equation}
\frac{d^2}{d\chi^2}\left(\chi \theta^b\right)=-2 \Phi_{,b}
\label{geoGR}
\end{equation}
where $,b$ indicates a partial derivative with respect to $\theta^b$, $\chi$ is the radial comoving distance, and $\chi\vec{\theta}=(\chi\theta^1, \chi\theta^2)$ is a two-component vector representing on-sky position.  This equation can be integrated to obtain the `true' on-sky position of the light source as a function of the observed on-sky position. The convergence is then given by taking the two-dimensional on-sky Laplacian ($\nabla^2$) of this expression:
\begin{equation}
\kappa_{GR}(\vec{\theta})=\frac{1}{2} \int_0^{\chi_\infty} d\chi\; \nabla^2
\Phi(\vec{\theta},\chi) g(\chi)
\label{kappagr}
\end{equation}
where $g(\chi)$ is the lensing kernel:
\begin{equation}
g(\chi)=2\chi\int_\chi^{\chi_\infty}d\chi^\prime\,\left(1-\frac{\chi}{\chi^\prime}\right)\,W\left(\chi^\prime\right),
\label{gdef}
\end{equation}
$W(\chi)$ is the normalised redshift distribution of the source galaxies, and $\chi_\infty$ is the comoving distance at $a\rightarrow 0$.

We compute the power spectrum of the convergence following closely the method laid out in \cite{DodelsonBook}. In the small angle approximation, it is straightforward to find:
\begin{align}
P^{i,j}_\kappa(l)&=\frac{1}{4}\int d^2 \theta\; e^{-i \vec{l} \cdot \vec{\theta}} \int_0 ^{\chi_{\infty}} d\chi \;g_i(\chi) \int_0 ^{\chi_{\infty}} d\chi' \;g_j(\chi') \nonumber \\ &\times \int \frac{d^3k}{(2\pi)^3}\; P_{\Phi}(k) k^4 e^{i \vec{k} \cdot [\vec{x}-\vec{x}']}
\label{pkappagr1}
\end{align}
where $\vec{x}$ labels three-dimensional position such that $\vec{x}=(\chi \theta_1, \chi \theta_2, \chi)$ and $\vec{x}'=(0, 0, \chi')$. $i$ and $j$ label the source redshift bins to be considered.

Performing the integrals over $\theta_1$ and $\theta_2$ and then over $k_1$ and $k_2$, we have:
\begin{align}
P_\kappa^{i,j}(l)&=\frac{1}{4}\int_0 ^{\chi_{\infty}} d\chi \;\frac{g_i(\chi)}{\chi^2} \int_0 ^{\chi_{\infty}} d\chi' \;g_j(\chi') \nonumber \\ &\times \int \frac{dk_3}{2\pi} \;P_{\Phi}\left(\sqrt{k_3^2+\frac{l^2}{\chi^2}}\right) k^4 e^{i k_3 \cdot [\chi-\chi']}.
\label{pkappagr2}
\end{align}
Finally, the Limber approximation \cite{Limber1953, Simon2007}, valid here on $l \gtrapprox 10$ \cite{Schmidt2008}, is employed, such that $k_3 \ll \frac{l}{\chi}$, and therefore
$k \approx \frac{l}{\chi}$.  The small angle limit also means that  $P^{i,j}_{\kappa}(l) \simeq P^{i,j}_{\kappa}(\ell)$, where $\ell$ labels an angular multipole \cite{Hu2000}. We find:
\begin{equation}
P^{i,j}_{\kappa}(\ell)=\frac{\ell^4}{4}\int_0^{\chi_\infty} d\chi\;
\frac{g_i(\chi)g_j(\chi)}{\chi^6}\, P_{\Phi}\left(\frac{\ell}{\chi}, \chi  \right).
\label{clGR_a}
\end{equation}
We have computed here the power spectrum of the convergence; that of the shear could be straightforwardly calculated by replacing equation \ref{kappagr} with the appropriate, similar definition.

\subsection{Calculating convergence: modified gravity}
\label{subsection:convergenceMG}

As indicated in equation \ref{defgammu}, generally in non-GR theories
$\Phi \neq \Psi$. So, in modified gravity equation \ref{geoGR} becomes:
\begin{equation}
\frac{d^2}{d\chi^2}\left(\chi \theta^b\right)=-(\Phi_{,b}+\Psi_{,b}).
\label{geoMG}
\end{equation}
The convergence then becomes:
\begin{align}
\kappa_{MG}&=\frac{1}{4} \int_0^{\chi_\infty} d\chi \;\nabla^2
\left[\Phi(\vec{\theta},\chi)+\Psi(\vec{\theta},\chi)\right] g(\chi)
\nonumber \\
&=\frac{1}{4} \int_{-\infty}^{0}dx\; \frac{c}{\mathcal{H}(x)} \nabla^2
\left[\Phi(\vec{\theta},x)+\Psi(\vec{\theta},x)\right] g(\chi(x))
\label{kappamg1}
\end{align}
where hereafter we will use $x=\mathrm{ln}(a)$ instead of $\chi$ or $a$, and we have converted the integration measure to $x$ using $d\chi = -c/\Hu\, dx$, where $\Hu=aH$ is the conformal Hubble factor.  Note that $x$ here is distinct from the three-dimensional position variable $\vec{x}$.

To calculate the power spectrum of the convergence under modifications to GR, we follow \cite{Baker2014} and perturb our field equations about those of the GR+$\Lambda$CDM model. Our reasoning here is that current observations only permit theories which can match GR+$\Lambda$CDM predictions to leading order; we are interested in determining next-to-leading order corrections that are still permitted. Note that we are building a theory of linear perturbations \textit{in model space}, which is distinct from \textit{spacetime} perturbation theory. We define the perturbations of the quasistatic functions $\mu$ and $\gamma$ about their GR values using:
\begin{align}
\mu(x, k)&=1+\delta \mu(x,k) \nonumber \\
\gamma(x,k)&=1+\delta \gamma(x,k). 
\label{defpert1}
\end{align}
In addition, we introduce a perturbation about the standard value of the effective equation of state of the non-matter sector, $w(x)$:
\begin{equation}
w(x)=-1+\beta(x),
\label{defbeta}
\end{equation}
and we define the useful related quantity:
\begin{equation}
u(x)=\int_0^x \beta(x') \,dx'.
\label{u}
\end{equation}

We now consider how these linear perturbation variables propagate through to $\kappa$ and hence
$P^{i,j}_\kappa(\ell)$.  Firstly, from equation \ref{defgammu}, we can write:
\begin{align}
\Phi(x,k)+\Psi(x,k)&=\left(1+\frac{1}{\gamma(x,k)}\right)\Phi(x,k) \nonumber \\
&\approx\left(2-\delta\gamma(x,k)\right)\Phi(x,k).
\label{psiphi}
\end{align}
In order to express our results as corrections to GR+$\Lambda$CDM, we need to relate $\Phi(x,k)$ to $\Phi_{GR}(x,k)$. There are two effects to be accounted for. Firstly, the relationship between $\Phi(x,k)$ and matter density perturbations can be altered. Secondly, if the field equations are modified, $\Delta_M(x,k)$ will evolve at a different rate, and hence will be displaced from its GR value. To account for this we introduce the deviation $\delta_\Delta(x,k)=\Delta_M(x,k)/\Delta_M^{GR}(x,k)-1$. 
In \cite{Baker2014} it was shown that $\delta_\Delta(x,k)$ is given by the following integral expression:
\begin{align}
\delta_{\Delta}(x,k)=\frac{3}{2}\int_{-\infty}^x\,\Omega^{GR}_M(\tilde{x})I(x,\tilde{x})\,\delta S_f(\tilde{x},k)\,d\tilde{x}.
\label{rty}
\end{align}
The integrand above separates into two parts: $\delta S_f(\tilde{x},k)$, which encapsulates all deviations from GR+$\Lambda$CDM, and $\Omega_M^{GR}(\tilde{x})I(x,\,\tilde{x})$, which is a weighting function containing GR+$\Lambda$CDM quantities only. It will be useful for us to present the explicit form of $\delta S_f(x,k)$ here, derived in \cite{Baker2014}: 
\begin{align}
\delta &S_f(x,k)=\delta \mu(x,k)-\delta\gamma(x,k) \nonumber \\ 
&+\frac{(1-\Omega_M^{GR})}{\Omega_M^{GR}}\Big[ 3\,\Omega^{GR}_M\,\left(1+f_{GR}(x)\right) u(x) +f_{GR}(x)\beta(x) \Big].
\label{deltaS_growth}
\end{align}
The explicit form of $I(x,\tilde{x})$ can be found in \cite{Baker2014}.

With these modifications in hand, the parameterised Poisson equation becomes:
\begin{align}
-2k^2\Phi(x,k)&=8\pi G e^{2x} \rho_M^{GR}(x) \Delta_M(x,k)\,(1+\delta \mu(x,k)) \nonumber \\
&=3{\cal H}_{GR}^2(x)\Omega_M^{GR}(x)\Delta^{GR}_M(x,k)\,\nonumber \\ &\times \left(1+\delta_\Delta(x,k)\right)\left(1+\delta\mu(x,k)\right)
\label{poissoneqn}
\end{align}
where in going from the first to the second line, we have used the fact that the combination $\Hu_{GR}^2(x)\Omega^{GR}_M(x)$ is unchanged by our modifications to the background expansion rate, as shown in Appendix \ref{appendix:H2Om}. Hence, $\Phi(x,k)$ is given in terms of $\Phi_{GR}(x,k)$ by:
\begin{align}
\Phi(x,k)&\simeq \Phi_{GR}(x,k)\left(1+\delta_\Delta(x,k)+\delta\mu(x,k)\right).
\label{poisson}
\end{align}

Combining equations \ref{psiphi} and \ref{poisson}, we now have an expression for $\Phi+\Psi$ in modified gravity in terms of the GR potential plus perturbative correction factors:
\begin{align}
\Phi(x,k)+\Psi(x,k) &\simeq\Phi_{GR}(x,k) \Big(2-\delta\gamma(x,k) \nonumber \\
&+2\delta_\Delta(x,k)+2\delta\mu(x,k)\Big).
\label{phipsi2}
\end{align}
So, referring to equation \ref{kappamg1}, $\kappa$ becomes:
\begin{align}
\kappa_{MG}(\vec{\theta})&=\frac{1}{4} \int_{-\infty}^{0}dx \frac{cg(\chi(x))}{\mathcal{H}(x)} \nabla^2
\Big[\Phi_{GR}(x,k) \nonumber \\ &\times (2+2\delta\mu(x,k)-\delta\gamma(x,k)+2\delta_\Delta(x,k))\Big].
\label{kappaMG2}
\end{align}
At this stage, it becomes more convenient to work directly with the power spectrum $P^{i,j}_\kappa(\ell)$.  This can be computed to linear order in deviations from GR+$\Lambda$CDM, in direct analogy to the method outlined for the GR case in Section \ref{subsection:convergenceGR}.  We find:
\begin{align}
P^{i,j}_\kappa(\ell)&=\frac{\ell^4}{4}\int_{-\infty}^{0} dx\frac{c}{\mathcal{H}(x)}
G_i(\chi(x)) G_j(\chi(x)) P_\Phi^{GR}\left(\frac{\ell}{\chi(x)} , \chi(x) \right) \nonumber \\ 
& \times   \left(1+2\delta\mu(x,k)-\delta\gamma(x,k)+2\delta_\Delta(x,k) \right)
\label{Pkappa1}
\end{align}
where we have defined $G_i(\chi(x))=\frac{g_i(\chi(x))}{\chi(x)^3}$.

There are still two non-GR effects to account for, both originating from the modified expansion history. If $\beta(x)\neq 0$ in equation \ref{defbeta}, $\Hu(x)$ and $\chi(x)$ will scale differently with the time variable $x$. Using the expression for $\delta\Hu(x)=\Hu(x)-\Hu_{GR}(x)$ derived in equation \ref{appeq1} in the Appendix, we find that 
\begin{align}
\frac{1}{\mathcal{H}(x)}&=\frac{1}{\mathcal{H}_{GR}(x)}\left(1-\frac{\delta \mathcal{H}}{\mathcal{H}_{GR}(x)}\right) \nonumber \\
&=\frac{1}{\mathcal{H}_{GR}(x)}\left( 1 - \frac{3}{2}u(x)(1-\Omega_M^{GR}(x))\right),
\label{delH}
\end{align}
and hence
\begin{align}
\chi(x)&\approx\int_x^0\frac{c}{\Hu_{GR}(x')}\left(1-\frac{\delta\Hu(x')}{\Hu_{GR}(x')}\right)\,dx' \nonumber \\
\Rightarrow \delta\chi(x)&\approx\frac{3}{2}\int_x^0\frac{c}{\Hu_{GR}(x')} u(x')
\left(1-\Omega_M^{GR}(x')\right)\,dx'
\label{delchi}
\end{align}
where $\delta \chi=\chi-\chi_{GR}$.

The deviation of $\chi(x)$ from its GR value will also affect quantities which depend on $\chi(x)$, such as $G(\chi(x))$ and $P^{GR}_\Phi(\ell/\chi(x))$ \footnote{Note that this remains true even though we have already related the power spectrum of the \textit{modified} potentials to $P_\Phi^{GR}$. Effectively, we have so far accounted for the modified perturbations, but not for the modified background expansion.}. We allow for this by expanding these in a Taylor series around $\chi_{GR}$, to first order:
\begin{align}
P^{GR}_\Phi\left(\frac{\ell}{\chi_{MG}}\right) &\approx P^{GR}_\Phi\left(\frac{\ell}{\chi_{GR}}\right)\left[1+\frac{\partial\,\mathrm{ln}P_\Phi}{\partial\,\mathrm{ln}\chi}\Bigg|_{\chi_{GR}}\frac{\delta\chi}{\chi_{GR}}\right]
\label{Pphidev} 
\end{align}
\begin{align}
 G_i(\chi)G_j(\chi)&\approx G_i(\chi_{GR})G_j(\chi_{GR})\nonumber \\ &\times \left[1 +\left(\frac{\partial\,\mathrm{ln}G_i(\chi)}{\partial \,\mathrm{ln}\chi}+\frac{\partial\,\mathrm{ln}G_j(\chi)}{\partial \,\mathrm{ln}\chi}\right)\Bigg|_{\chi_{GR}}\frac{\delta\chi}{\chi_{GR}}\right]
\label{Gdev}
\end{align}
where $\delta\chi$ is given by equation \ref{delchi} above. We have now accounted for all modified gravity effects, and these are summarised in Table \ref{table:MG_Effects}.  

\begin{table*}[t]
{\renewcommand{\arraystretch}{2.7}
\begin{tabular}{ccc}\hline  \hline
{\bf Correction} & {\bf Description} & {\bf Equation} \\ \hline
$\Phi(x,k) + \Psi(x,k) \simeq (1+\frac{1}{\gamma(x,k)})\Phi(x,k)$ & Non-unity ratio of scalar
potentials & \ref{psiphi} \\ %\hline
$\Phi(x,k)\simeq\Phi_{GR}(x,k)(1+\delta_{\Delta}(x,k)+\delta \mu(x,k))$ & Altered Poisson
equation  & \ref{poisson} \\ %\hline
$\frac{1}{\mathcal{H}(x)} \simeq \frac{1}{\mathcal{H}_{GR}(x)}\left[ 1 -
  \frac{3}{2}u(x)(1-\Omega_M^{GR}(x))\right]$ & Altered
$\mathcal{H}(x)$ & \ref{delH} \\
%\hline
$\chi(x)\simeq\chi_{GR}(x)+\frac{3}{2}\int\frac{c}{\Hu_{GR}(x)} u(x)
\left(1-\Omega_M^{GR}(x)\right)\,dx$ & Altered $\chi$  & \ref{delchi} \\ %\hline
$ G_i(\chi)G_j(\chi)\simeq G_i(\chi_{GR})G_j(\chi_{GR})\left[1 +\left(\frac{\partial\,\mathrm{ln}G_i(\chi)}{\partial \,\mathrm{ln}\chi}+\frac{\partial\,\mathrm{ln}G_j(\chi)}{\partial \,\mathrm{ln}\chi}\right)\Bigg|_{\chi_{GR}}\frac{\delta\chi}{\chi_{GR}}\right]$
& Altered $G(\chi)$  &
\ref{Gdev} \\ %\hline
$P^{GR}_\Phi\left(\frac{\ell}{\chi}\right) \simeq P^{GR}_\Phi\left(\frac{\ell}{\chi_{GR}}\right)\times\left[1-\frac{\partial \ln (k^{-4}\,P^{GR}_{\delta}(x=0,k))}{\partial
    \ln k}\Bigg|_{k=\ell/\chi_{GR}}\frac{\delta\chi}{\chi_{GR}}\right]$ &
Altered $P^{GR}_\Phi$  &
\ref{Pphideriv} \\ \hline \hline
\end{tabular} } 
\caption{Here we summarise the various corrections to the GR expression for $P_\kappa^{i,j}(\ell)$, including a brief description and the number of the equation in which they are introduced.}
\label{table:MG_Effects}
\end{table*}

Finally, it will be more convenient for us to work in terms of $P_\delta$, the matter power spectrum, instead of $P_\Phi$.  We do so via the following expression, where for clarity we temporarily omit the label `GR' on all quantities.:
\begin{align}
P_\Phi\left(k, x\right)=\frac{1}{k^4}\frac{9}{4}\left(\frac{\mathcal{H}(x)}{c}\right)^4\Omega_M^2(x) D(x)^2P_{\delta}\left(x=0, k\right).
\label{PphitoPdel}
\end{align}
Here $D(x)$ is the usual growth factor of matter perturbations. Inserting equation \ref{PphitoPdel} into equation \ref{Pphidev}, we find:
\begin{align}
P^{GR}_\Phi\left(\frac{\ell}{\chi}\right) &\approx P^{GR}_\Phi\left(\frac{\ell}{\chi_{GR}}\right)\times\nonumber\\
&\left[1-\frac{\partial \ln (k^{-4}\,P^{GR}_{\delta}(x=0, k))}{\partial \ln k}  \Bigg|_{k=\ell/\chi_{GR}}\frac{\delta\chi}{\chi_{GR}}\right].
\label{Pphideriv}
\end{align}
Drawing together, then, equations \ref{Pkappa1}, \ref{delH}, \ref{Gdev} and \ref{Pphideriv}, and using equation \ref{PphitoPdel}, we obtain our final expression for the convergence power spectrum under modifications to general relativity:
\begin{widetext}
\begin{align}
&P^{i,j}_\kappa(\ell)=\frac{9}{16}\int_{-\infty}^{0}dx\;\frac{g_i(\chi_{GR}(x))g_j(\chi_{GR}(x))}{\chi_{GR}(x)^2}P_{\delta}^{{GR}}\left(\frac{\ell}{\chi_{GR} (x)}\right)D^2_{GR}(x)\frac{\Hu_{GR}^3(x)}{c^3}\Omega_M^{GR}(x)^2\times \Bigg[1+\frac{3}{2}u(x)\left(1-\Omega_M^{GR}(x)\right) \nonumber \\ &+2\delta\mu(x,k)-\delta\gamma(x,k)+2\delta_\Delta(x,k)+\left(\frac{\partial \mathrm{ln}G_i(\chi)}{\partial \mathrm{ln}\chi}+\frac{\partial \mathrm{ln}G_j(\chi)}{\partial  \mathrm{ln}\chi}-\frac{\partial \mathrm{ln}(P^{GR}_{\delta}(x=0, k)/k^4)}{\partial \mathrm{ln}k}\right)\Bigg|_{\chi_{GR}}\frac{\delta\chi(x)}{\chi_{GR}(x)}\Bigg].
\label{Pkappafinal}    
\end{align}
\end{widetext}
The major advantage of equation \ref{Pkappafinal} is that it neatly separates the convergence power spectrum into the familiar GR expression (the non-bracketed quantity) and a correction factor (the bracketed terms). It is then easy to pick out contributions from:
\begin{itemize}
\item the modified clustering properties (described by $\delta\mu$ and $\delta\gamma$),
\item the modified expansion history (described by $\beta$, $u$ and $\delta\chi$), and
\item the modified growth rate of matter density perturbations (encapsulated in $\delta_\Delta$, see equation \ref{rty}).
\end{itemize}

It will be useful for us to write equation \ref{Pkappafinal} in a form
which explicitly highlights the GR expression and
the correction factor:
\begin{equation}
P_\kappa^{i,j}(\ell)=\int_{-\infty}^{0} dx\;
\mathcal{K}(x,\ell) \big(1+\delta S_{WL}(x,\ell)\big).
\label{deltaCl}
\end{equation}
Here we have defined the `kernel' term:
\begin{align}
\mathcal{K}(x,\ell)&=\frac{9}{16}\frac{g_i(\chi_{GR}(x))g_j(\chi_{GR}(x)}{\chi_{GR}(x)^2}P_\delta^{GR}\left(\frac{\ell}{\chi_{GR}(x)}\right)
\nonumber \\ & \times D_{GR}^2(x)
\frac{\mathcal{H}_{GR}^3(x)}{c^3}\Omega_M^{GR}(x)^2,
\label{K}
\end{align}
and the `source' term:
\begin{align}
&\delta S_{WL}(x,\ell)= \frac{3}{2}u(x)\left(1-\Omega_M^{GR}(x) \right) 
\nonumber \\ &+2 \delta \mu(x,k) -\delta \gamma(x,k)  +2 \delta_ \Delta (x,k)+ \Bigg(\frac{\partial\,\mathrm{ln}G_i(\chi)}{\partial \,\mathrm{ln}\chi} \nonumber \\ &+\frac{\partial\,\mathrm{ln}G_j(\chi)}{\partial \,\mathrm{ln}\chi}-
  \frac{\partial \mathrm{ln} [P^{GR}_\delta(x=0,k)/k^4]}{\partial \mathrm{ln}k}
\Bigg)\Bigg|_{\chi_{GR}}\frac{\delta \chi(x)}{\chi_{GR}(x)}.
\label{deltaS}
\end{align}

\section{Understanding degeneracies with the linear response approach} 
\label{section:under_deg}
\noindent 
We have at hand an expression (equation \ref{Pkappafinal}) for $P_\kappa^{i,j}(\ell)$ under modifications to general relativity. Let us now investigate what this can teach us about the degeneracies between gravitational parameters in weak lensing observations. Note that we restrict ourselves to discussing degeneracies between parameters describing modifications to gravity. We do not examine degeneracies between gravitational and cosmological parameters, nor do we investigate degeneracies with the galaxy bias. We leave these questions for future work.

In this section, we consider the case in which $\delta \mu$ and $\delta \gamma$ are independent of scale, due to the fact that the scale-dependence of these functions is expected to be sub-dominant to their time-dependence \cite{BakerScales2014, Silvestri2013, Hojjati2014}. We will briefly investigate scale-dependence later, in Section \ref{subsection:scaledep_forecasts}. Additionally, as we are working in the quasistatic approximation, our analysis is restricted to the regime of validity of linear cosmological perturbation theory.  Various values of $\ell_{max}$ which ensure this to be true are suggested in the literature (see for example \cite{Schmidt2008,Thomas2009}). Adopting a conservative approach, we select $\ell_{max}=100$ here and for the remainder of this work.

We first remind the reader of how degeneracy directions may be calculated. Then, using equations \ref{deltaS_growth} and \ref{Pkappafinal}, we explore how the degeneracy directions of weak lensing and redshift-space distortions in the space of the parameters of $\delta \mu(x)$ and $\delta \gamma(x)$ are affected by the chosen ansatzes for these functions. Note that here and for the remainder of this work, we compute the GR+$\Lambda$CDM matter power spectrum using the publicly available code CAMB \cite{Lewis2000} and using the best-fit $\Lambda$CDM parameters of the 2013 Planck release (including Planck lensing data) \cite{Planck2013}.

\subsection{Calculating degeneracy directions}
\label{subsection:deg_dir}
\noindent
Degeneracies exist when an observation can probe only some combination of the parameters we wish to constrain. The degeneracy direction is the relationship between parameters in the fiducial scenario (here, in GR+$\Lambda$CDM). For example, if this relationship is $a=b$, then the relevant observation can probe only $a-b$, not $a$ or $b$ individually.  

In the case of weak lensing, degeneracy directions can be understood in the following schematic way.  First, define the fractional difference between $P^{i,j}_\kappa(\ell)$ in an alternative gravity theory and in GR+$\Lambda$CDM:
\begin{equation}
\delta P^{i,j}_\kappa(\ell)=\frac{P^{i,j}_\kappa(\ell)-P^{i,j}_{\kappa,GR}(\ell)}{P^{i,j}_{\kappa,GR}(\ell)}.
\label{deltapk}
\end{equation}
To find the degeneracy direction, we find the relationship which exists between parameters when $\delta P^{i,j}_\kappa(\ell)=0$.  If we consider a two parameter case (call them $a$ and $b$), straightforward algebra allows us to find an expression of the form
\begin{equation}
a=\mathcal{D}(\ell)b
\label{degexample}
\end{equation}  
where $\mathcal{D}(\ell)$ may be a complicated expression, but depends only on GR+$\Lambda$CDM quantities.  The degeneracy direction, we see, depends on $\ell$ in the weak lensing case. 

In order to calculate $\delta P_\kappa^{i,j}(\ell)$, we need to specify $W_i$, the normalised redshift distribution of the source galaxies in the redshift bin $i$.  We select a source number density with the following form:
\begin{equation}
n(z) \propto z^{\alpha} e^{-\left(\frac{z}{z_0}\right)^\beta},
\label{nofz}
\end{equation}
and we select $\alpha=2$, $\beta=1.5$, and $z_0=z_m/1.412$ where $z_m=0.9$ is the median redshift of the survey, mimicking the number density of a Dark Energy Task Force 4 type survey \cite{Thomas2009,Amendola2008}.  In this section, we will simply consider all galaxies between $z=0.5$ and $z=2.0$ to be in a single redshift bin, with $W(\chi)$ given by normalising equation \ref{nofz}.  

To break parameter degeneracy in a two parameter case, a second observable with a different (ideally orthogonal) degeneracy direction is introduced.  Here, we choose this second observable to be redshift-space distortions, as it is known to provide nearly orthogonal constraints to weak lensing. We will therefore often employ results from \cite{Baker2014}.  Particularly, we reproduce here their equation for the deviation of $f\sigma_8(x)$ from its GR value, analogous to our equation \ref{deltapk}:
\begin{align}
\delta f\sigma_8(x)&=\frac{f\sigma_8(x)-f\sigma_8^{GR}(x)}{f\sigma_8^{GR}(x)}=\int_{-\infty}^{x} G_f(x,\tilde{x})\delta S_f(\tilde{x}) d\tilde{x}
\label{deltafsig8}
\end{align}
where $\delta S_f(x)$ is given as in our equation \ref{deltaS_growth}, and $G_f(x,\tilde{x})$ is a general relativistic kernel given in equation 34 of \cite{Baker2014}. Note that $f\sigma_8(x)$ above is independent of $k$, because we are considering a case where $\mu$ and $\gamma$ are functions of time only.

The degeneracy direction of a measurement of $f\sigma_8(x)$ can then be computed in a directly analogous way to that described above for weak lensing.  The sole difference is that instead of depending on multipole $\ell$, the degeneracy direction is dependent on the time of observation, $x$.  

With this information in hand, we now explore degeneracy directions of weak lensing and redshift-space distortions in the space of the parameters of $\delta \mu(x)$ and $\delta \gamma(x)$.  

\subsection{Degeneracy directions in the $\bar{\mu}_0-\Sigma_0$ plane}
\label{subsection:betazero_deg:}
\noindent
As mentioned above, redshift-space distortions are the preferred choice of an additional observation to break weak lensing degeneracy in this scenario. Upon closer examination, this statement hinges upon the chosen time-dependent ansatz for the functions which parameterise deviations from GR. As there is no clear front-runner amongst alternative theories of gravity, typically a phenomenological ansatz is chosen, in which deviations from GR become manifest at late times in order to mimic accelerated expansion. It is for this type of phenomenological ansatz that redshift-space distortion and weak lensing observations are known to provide complimentary constraints \cite{Simpson2012}.

However, it may also be desirable to constrain the parameters of a specific theory of gravity. The functions which parameterise the deviation of an alternative gravity theory from GR can, in principle, take on a wide range of time-dependencies. Is the combination of weak lensing and redshift-space distortions still an effective way to break degeneracies and constrain the parameters of the theory we consider?  A priori, this is unknown.  

To explore this issue, we consider now the degeneracy directions of weak lensing and redshift-space distortions under two different ansatzes for the functions which parameterise deviations from GR. For this section only, we make the simplifying assumption that $\beta(x)=0$ (i.e. the expansion history is $\Lambda$CDM-like). We expect that the effect of this assumption on our qualitative findings will be small. 

First, we perform a simple operation on $\delta \mu(x)$ and $\delta \gamma(x)$ to obtain a more observationally-motivated set of functions.  Let us call these $\bar{\mu}(x)$ and $\Sigma(x)$, in keeping insofar as possible with the notation used in \cite{Simpson2012}. The choice of this set of functions allows nearly orthogonal constraints in the $\bar{\mu}_0-\Sigma_0$ plane for the phenomenological choice of time-dependence. The mapping between the two sets of functions, as shown in Appendix \ref{appendix:convertfuncs}, is given by:
\begin{align}
\Sigma(x)&=\delta \mu(x)- \frac{1}{2}\delta \gamma(x) \nonumber \\
\bar{\mu}(x)&= \delta \mu(x)- \delta \gamma(x).
\label{param2}
\end{align}
We can rewrite the linear response `source' terms for both weak lensing (equation \ref{deltaS}) and redshift-space distortions (equation \ref{deltafsig8}) in terms of $\bar{\mu}(x)$ and $\Sigma(x)$ (in the $\beta(x)=0$ case):
\begin{align}
\delta S_{WL} (x)&=2\Sigma(x)  + 3\int_{-\infty}^x\,\Omega^{GR}_M(\tilde{x})I(x,\tilde{x})\,\bar{\mu}(\tilde{x}) d\tilde{x} \nonumber \\
\delta S_{f}(x) &= \bar{\mu}(x).
\label{deltaSbetazero2}
\end{align}
We see that $\delta S_{f}(x)$ depends solely on $\bar{\mu}(x)$. 

The expression for $\delta S_{WL}(x)$ requires slightly more pause.  It depends on $\Sigma(x)$, but it also contains another term, which comprises an integral over $\bar{\mu}(x)$ and some general relativistic quantities. By comparing with equation \ref{rty}, we can easily recognise this term as $2 \delta_\Delta(x)$.  This term quantifies a correction to the degeneracy direction of weak lensing away from $\Sigma_0 =0$. It is clearly dependent upon the ansatz of time-dependence chosen for $\bar{\mu}(x)$.  Particularly, we note that due to the integral nature of the correction term, choices of $\bar{\mu}(x)$ which persist significantly over longer times will result in greater deviations to the degeneracy direction. 

We now consider two ansatzes for $\bar{\mu}(x)$ and $\Sigma(x)$. First, consider a phenomenological ansatz, for which we know weak lensing and redshift-space distortions to be an effective combination in constraining gravity theories.  This choice is a specific case of the form proposed in \cite{Ferreira2010} and has been used in, for example, \cite{Simpson2012}. It is given by:
\begin{align}
\bar{\mu}(x)&= \bar{\mu}_0\frac{\Omega^{GR}_{\Lambda}(x)}{\Omega^{GR}_{\Lambda}(x=0)} \nonumber \\ 
\Sigma(x)&= \Sigma_0\frac{\Omega^{GR}_{\Lambda}(x)}{\Omega^{GR}_{\Lambda}(x=0)}
\label{latetimemuSig}
\end{align}
where $\Omega_\Lambda^{GR}(x)$ is the time-dependent energy density of dark energy in the fiducial $\Lambda$CDM cosmology.  

We insert $\delta S_{WL}(x)$ (equation \ref{deltaSbetazero2}) into equations \ref{deltapk} and \ref{deltafsig8} with our chosen $\bar{\mu}(x)$ and $\Sigma(x)$.  We then follow the procedure sketched in Section \ref{subsection:deg_dir} to find the degeneracy directions of weak lensing and redshift-space distortion in the $\bar{\mu}_0-\Sigma_0$ plane.  In this particular case the degeneracy direction of redshift-space distortions does not depend on time.  This is because $\delta S_f(x)$ is dependent on only one parameter, $\bar{\mu}_0$, and therefore the only degeneracy direction is $\bar{\mu}_0=0$.  

\begin{figure*}[t]
\begin{center}
\subfigure{\label{figure:under_OmL}\includegraphics[width=0.48\linewidth]{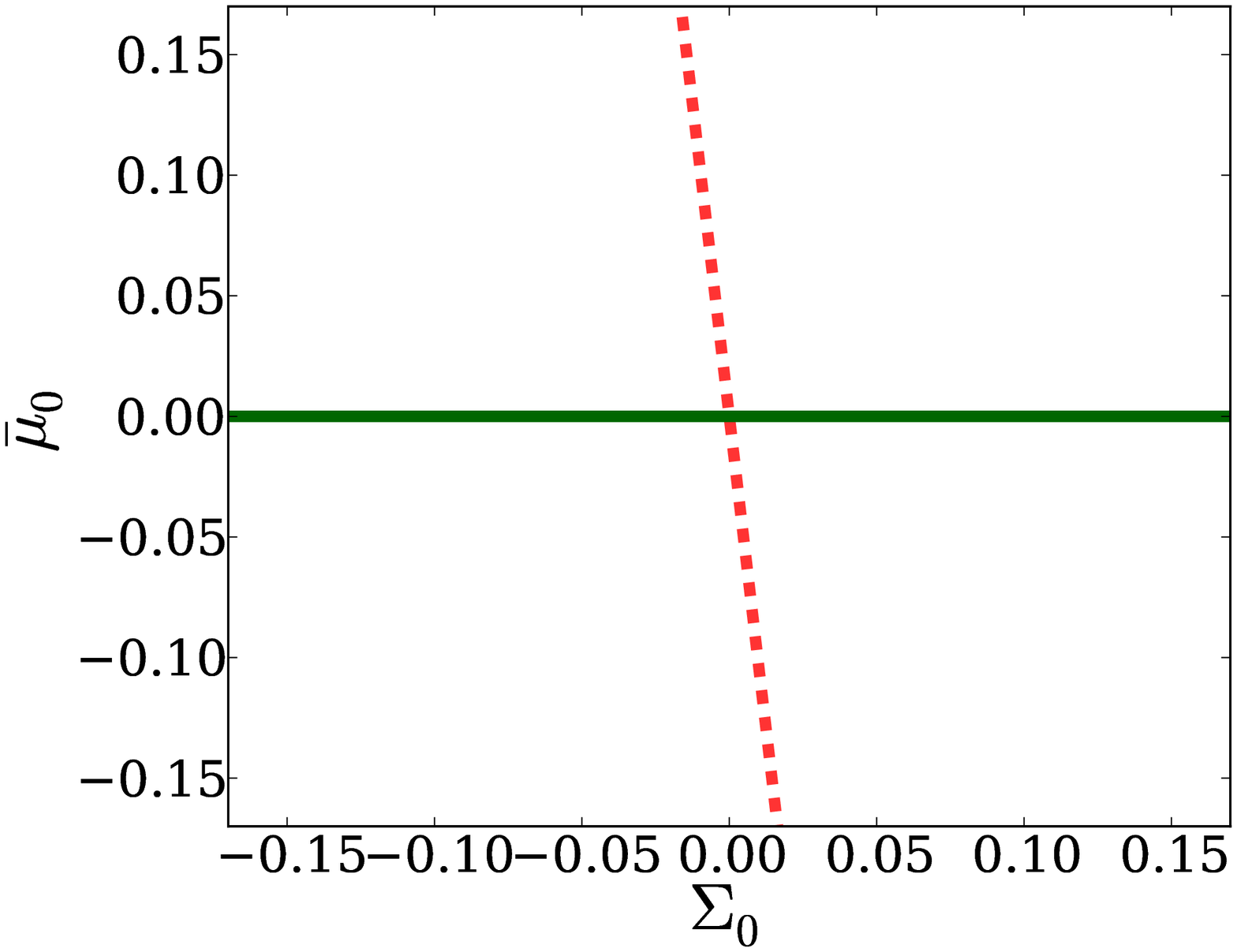}}
\subfigure{\label{figure:under_const}\includegraphics[width=0.48\linewidth]{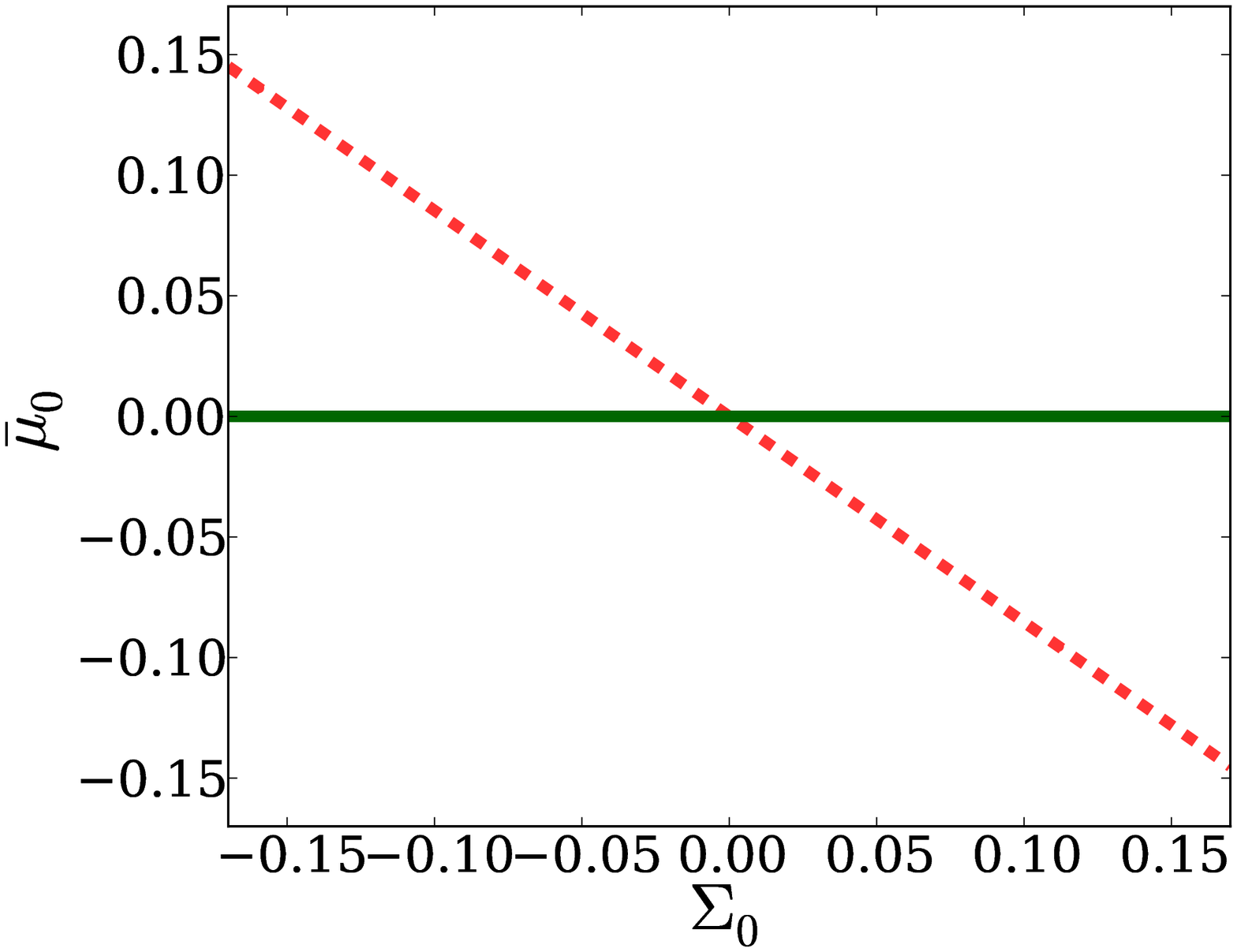}}
\end{center}
\caption{Degeneracy directions of weak lensing ($\ell=50$, dashed red) and redshift-space distortions (solid green) in the $\bar{\mu}_0-\Sigma_0$ plane, where $\bar{\mu}(x)$ and $\Sigma(x)$ scale as $\Omega_\Lambda^{GR}(x)$ (left), and as constants (right).}
\label{figure:degeneracydirections_betazero}
\end{figure*}

The degeneracy directions for this ansatz can be seen in Figure \ref{figure:degeneracydirections_betazero} (left).  In the case of weak lensing, we have plotted the degeneracy direction for $\ell=50$; directions for other multipoles $\ell=10-100$ differ only within 5\%. We see that, indeed, the degeneracy directions are nearly orthogonal, with only a slight correction of the weak lensing degeneracy direction away from $\Sigma_0=0$.  

Now, consider selecting an ansatz with a very different time-dependence. To guide our selection, recall that we expect choices of $\bar{\mu}(x)$ which persist over longer times to result in a greater value of the integral term in equation \ref{deltaSbetazero2}, and hence a greater deviation of the weak lensing degeneracy direction from $\Sigma_0=0$.  Therefore, with no attempt to correspond to any particular gravity theory, we select the simplest possible choice which persists over long times: constant $\bar{\mu}(x)$ and $\Sigma(x)$.
\begin{align}
\Sigma(x)&=\Sigma_0 \nonumber \\ 
\bar{\mu}(x)&=\bar{\mu}_0.
\label{SigmuConst}
\end{align}
In reality, we use step functions beginning at $z=15$ rather than true constants to allow for the numerical computation of the degeneracy directions. The degeneracy directions are calculated as before, and are plotted in Figure \ref{figure:degeneracydirections_betazero} (right).  Clearly, they are less orthogonal than in the previous case, as expected from the comments above.

What does this example tell us about the effectiveness of combining weak lensing and redshift-space distortions? The ansatz for $\bar{\mu}(x)$ and $\Sigma(x)$ given by equation \ref{SigmuConst} deviates from GR+$\Lambda$CDM at all times after $z=15$. As mentioned above, most cosmologically-motivated alternative theories of gravity present deviations from GR+$\Lambda$CDM at late times only, mimicking accelerated expansion. Therefore, we treat the case of equation \ref{SigmuConst} as a heuristic `upper bound' on the cumulative effect produced by the integral term of equation \ref{deltaSbetazero2}. The effect of this term can be quantified by considering the angle of the weak lensing degeneracy direction with respect to the vertical. We find that for the range of $\ell$ which we consider and for the ansatz given by equation \ref{SigmuConst}, the maximum possible value of this angle is $\theta \approx 50^{\circ}$.  Although the degeneracy directions in this case are certainly no longer orthogonal ($\theta= 0^{\circ}$), they are sufficiently distinct that we expect the resulting constraints to be reasonable (if not ideal). We have therefore shown that the effectiveness of combining weak lensing with redshift-space distortions in the $\beta(x)=0$ case is relatively robust to the chosen form of $\bar{\mu}(x)$ and $\Sigma(x)$.

\section{Forecast constraints from future surveys}
\label{section:constraints}
\noindent
In addition to providing an understanding of degeneracies, our expression for $P^{i,j}_\kappa(\ell)$ enables the forecasting of constraints.  The straightforward form of equation \ref{deltaCl} renders the calculation of Fisher matrices very simple, and clarifies the interpretation of the resulting forecasts. We take advantage of these features to forecast constraints on gravitational parameters for a Dark Energy Task Force 4 (DETF4) type survey, as defined in the classification of \cite{Albrecht2006}. We focus on combined constraints from weak lensing and redshift-space distortions, with some consideration given as well to baryon acoustic oscillations.

As mentioned above, the forecasts presented here employ the technique of Fisher forecasting (see, for example, \cite{Bassett2011}).  The key quantity of this method is the Fisher information matrix:  
\begin{equation}
\mathcal{F}_{ab}=-\Big\langle \frac{ \partial^2 \ln \mathcal{L}}{\partial p_a \partial p_b } \Big\rangle
\label{Fisherdef}
\end{equation}
where ${p_i}$ are the relevant parameters, and $\mathcal{L}$ is the likelihood.  For redshift-space distortions, we straightforwardly build on the results of \cite{Baker2014} to construct the appropriate Fisher matrix. However, for weak lensing we require a slightly different expression. Although our equation \ref{Pkappafinal} allows for the cross-correlation of source galaxy redshift bins, we have until now considered only a single wide redshift bin. In practice, weak lensing data are normally considered in a number of tomographic redshift bins.  In \cite{Hu1999}, the Fisher matrix for such a situation is shown to be given by:
\begin{equation}
\mathcal{F}_{ab}=\sum_{\ell=\ell_{min}}^{\ell_{max}} \left(\ell+\frac{1}{2}\right)f_{sky}\mathrm{Tr}\left[C_{GR}^{-1}C_{,a}C_{GR}^{-1}C_{,b}\right]
\label{Fishertomo}
\end{equation}
where $,a$ is a derivative with respect to $p_a$, $f_{sky}$ is related to the fraction of the sky observed ($f_{sky}=0.375$ for a DETF4-type survey), and $C$ is an $N_b \times N_b$ matrix where $N_b$ is the number of tomographic redshift bins.  $C$ represents the observed power spectrum of the convergence, and is given by the following expression \cite{Hu1999}:
\begin{equation}
C^{i,j}(\ell)=P_{\kappa}^{i,j}(\ell)+\frac{\langle \gamma_{int}^2 \rangle \delta_{ij}}{\bar{n}_i}
\label{cij}
\end{equation}
where $\bar{n}_i$ is the number density of galaxies per steradian in bin $i$ and $\langle \gamma_{int}^2 \rangle$ is the rms intrinsic shear, equal to $0.22$ for a DETF4-type survey.

Computing the appropriate value of $\bar{n}_i$ requires the selection of source redshift bins.  In practice, once in possession of data, the selected bins are those which are maximal in number while maintaining shot noise sufficiently below the signal.  For our forecasting purposes, we instead follow, for example, \cite{Thomas2009} and \cite{Amendola2008}.  We select redshift bins by subdividing $n(z)$ of equation \ref{nofz} into 5 sectors, such that the number of galaxies in each bin is equal.  The value of $\bar{n}$ for the total redshift range for a DETF4-type survey is given by $\bar{n}=3.55 \times 10^8$, so the value in each tomographic bin is simply $\bar{n}_i=\bar{n}/5$.   

In the following subsections, we use the Fisher formalism to compute forecast constraints in a number of scenarios. We first consider constraints on the parameters of $\delta \mu(x)$ and $\delta \gamma(x)$ in the case where we fix the expansion history to mimic $\Lambda$CDM.  We then incorporate expected measurements of $w_0$ and $w_a$ from Baryon Acoustic Oscillations to forecast constraints on the parameters of $\delta \mu(x)$ and $\delta \gamma(x)$ in the case where we marginalise over the parameters of $\beta(x)$. We finish by discussing the directions of best constraint in the parameter space of the scale dependent ansatz for $\mu(x,k)$ and $\gamma(x,k)$ put forth in \cite{Silvestri2013}.

\subsection{$\Lambda$CDM-like expansion history: $\beta(x)=0$}
\label{subsection:betazero_forecasts}
\noindent
We first consider constraints on the parameters of $\delta \mu(x)$ and $\delta \gamma(x)$ in the case where the expansion history is fixed to be $\Lambda$CDM-like. As in equation \ref{param2}, we transform $\delta \mu(x)$ and $\delta \gamma(x)$ to $\bar{\mu}(x)$ and $\Sigma(x)$, and we choose the time-dependence given by equation \ref{latetimemuSig}.   

To compute these constraints, we calculate the $2 \times 2$ Fisher matrix for lensing, for redshift-space distortions, and for both observations combined. For this, we require expressions for the derivatives $\frac{\partial f\sigma_8(x)}{\partial \bar{\mu}_0}$, $\frac{\partial f\sigma_8(x)}{\partial \bar\Sigma_0}$ and
\begin{align}
\frac{\partial}{\partial \bar{\mu}_0} \left(P^{i,j}_\kappa(\ell)+\frac{\langle \gamma_{int}^2 \rangle \delta_{ij}}{\bar{n}_i} \right) &=\frac{\partial P^{i,j}_\kappa(\ell)}{\partial \bar{\mu}_0}\\
 \frac{\partial}{\partial \Sigma_0} \left(P^{i,j}_\kappa(\ell)+\frac{\langle \gamma_{int}^2 \rangle \delta_{ij}}{\bar{n}_i} \right)&=\frac{\partial P^{i,j}_\kappa(\ell)}{\partial \Sigma_0}
 \end{align}
  These are found in a straightforward manner from equations \ref{Pkappafinal} and \ref{deltafsig8}; we present them in Appendix \ref{appendix:pderivs}. 
  
\begin{figure}[ht]
\begin{center}
\includegraphics[width=\linewidth]{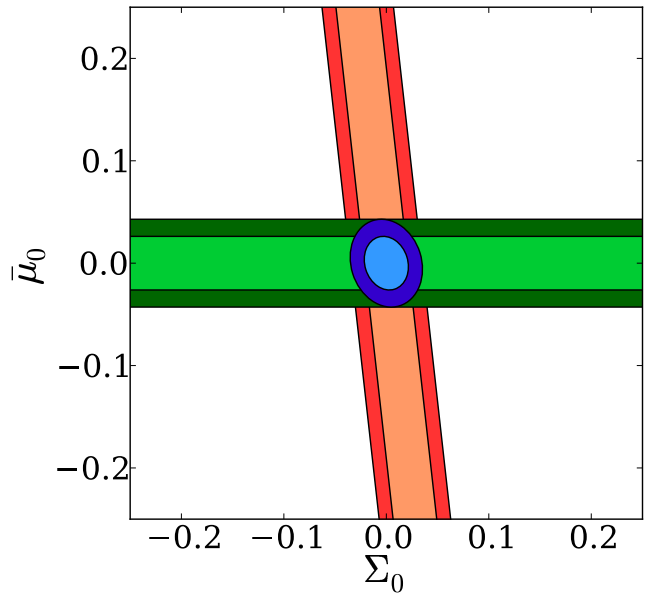}
\caption{Forecast constraints for weak lensing (orange), redshift-space distortions (green) and both observables combined (blue) for a DETF4-type survey, in the $\bar{\mu}_0-\Sigma_0$ plane with $\beta(x)$ fixed to $0$.  Contours represent the $68.3\%$ and $95.4\%$ confidence regions.}
\label{figure:forecastsmuSigbeta}
\end{center}
\end{figure}

The resulting forecast constraints are illustrated in Figure \ref{figure:forecastsmuSigbeta}.  As discussed in Section \ref{section:under_deg}, the degeneracy directions of the two observables are nearly orthogonal in this case.  Combining them results in promising forecast constraints on $\bar{\mu}_0$ and $\Sigma_0$.  We see that we can expect a DETF4-type survey to provide constraints at a level of approximately $4\%$ in this plane, in the case where $\beta(x)$ is assumed to be fixed at $0$. 

\subsection{The effect of marginalising over $\{w_0, w_a\}$}
\label{subsection:bao_forecasts}
\noindent
In reality, $\beta(x)$ is not fixed to zero, but rather the associated parameters will also be constrained with some non-zero error. While weak lensing and redshift-space distortions will provide some constraints on these, it is baryon acoustic oscillation (BAO) measurements which are expected to provide the best constraints on the expansion history of the universe. 

In this section, we use a CPL-type ansatz for $\beta(x)$ as proposed in \cite{Chevallier2001, Linder2003}: $\beta(x)=w_0 + 1 + w_a (1 - e^x)$. We incorporate forecast BAO constraints on $w_0$ and $w_a$ and use these to obtain expected constraints in the $\bar{\mu}_0-\Sigma_0$ plane. We first marginalise over only $w_0$, while holding $w_a$ to its fiducial value of $0$; then we examine the effect of allowing $w_a$ to vary as well.

\subsubsection{Marginalising over $w_0$; $w_a=0$}
\label{subsubsection:wafixed}
\noindent
We first demonstrate how constraints in the $\bar{\mu}_0-\Sigma_0$ plane are affected by marginalising over $w_0$ when $w_a$ is held fixed to its fiducial value of $0$.  

Because we are now incorporating information about three parameters ($\bar{\mu}_0$, $\Sigma_0$ and $w_0$), our Fisher matrices are $3 \times 3$ in dimension. In order to compute these, we now require additional derivatives of $P_\kappa^{i,j}(\ell)$ and $f\sigma_8(x)$ with respect to $w_0$; all are listed in Appendix \ref{appendix:pderivs}. Because transverse measurements of BAO are independent of non-background gravitational effects \cite{Bassett2010}, the Fisher matrix of BAO is non-zero only in the $(w_0, w_0)$ component. The value of this matrix component is equal to $\frac{1}{\sigma_{w_0, BAO}^2}$, where $\sigma_{w_0, BAO}$ is the $1$-$\sigma$ error on $w_0$ from BAO measurements.  

\begin{figure*}[t]
\begin{center}
\includegraphics[width=0.65\linewidth]{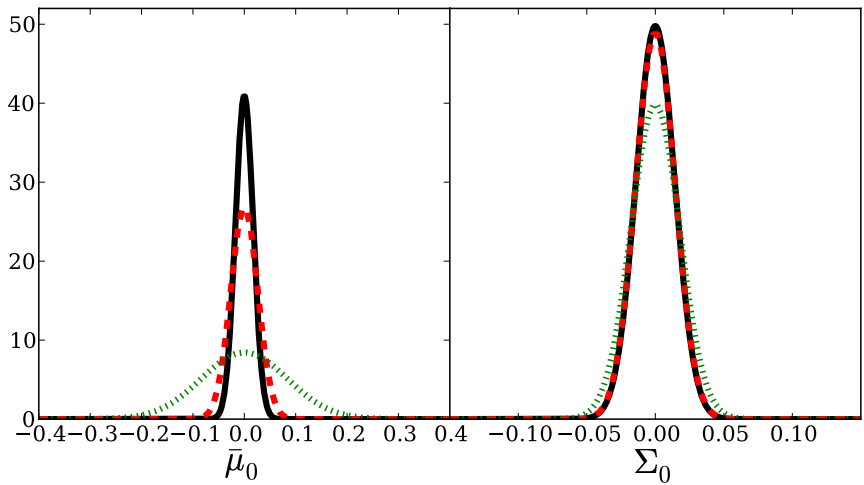}
\end{center}
\caption{Forecast constraints from weak lensing, redshift-space distortions, and BAO in the case where $w_0$ has been marginalised over and $w_a$ has been fixed to $0$. The left-hand panel shows the confidence region for $\bar{\mu}_0$ when $\Sigma_0$ is marginalised over, while the right-hand panel shows the confidence region for $\Sigma_0$ with $\bar{\mu}_0$ marginalised over. Black, solid: $w_0$ fixed; red, dashed: BAO error on $w_0=1 \%$ (DETF4); green, dotted: BAO error on $w_0=5\%$.}
\label{figure:withbao_1D}
\end{figure*}

\begin{figure}[ht]
\begin{center}
\includegraphics[width=\linewidth]{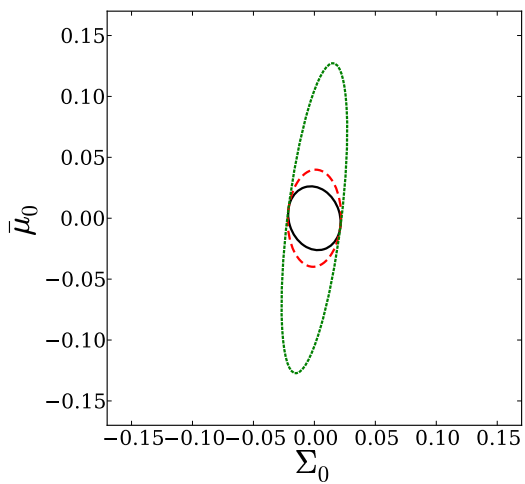}
\caption{Forecast $68.3\%$ confidence regions in the $\bar{\mu}_0-\Sigma_0$ plane, marginalising over $w_0$, for the case where $w_a=0$. Black, solid: $w_0$ fixed; red, dashed: BAO error on $w_0=1 \%$ (DETF4); green, dotted: BAO error on $w_0=5\%$.}
\label{figure:withbao_2D}
\end{center}
\end{figure}

To explore the effect of marginalising over $w_0$, we consider three levels of constraint from BAO:
\begin{enumerate}
\item{For comparison: the case where $w_0$ is fixed to its fiducial value.  This is identical to the case considered in Section \ref{subsection:betazero_forecasts}.}
\item{The case where $\sigma_{w_0, BAO}=1\%$. This scenario mimics best-case constraints from a DETF4-type survey.}
\item{The case where $\sigma_{w_0, BAO}=5\%$. This lies between current best constraints and scenario $2$ above.}
\end{enumerate}
The resulting constraints from the combination of weak lensing, BAO, and redshift-space distortions are shown in Figure \ref{figure:withbao_1D} and Figure \ref{figure:withbao_2D}. Figure \ref{figure:withbao_1D} shows in the left-hand panel the forecast constraints on $\bar{\mu}_0$ for cases $1-3$ above when marginalising over $w_0$ and $\Sigma_0$; the right-hand panel displays the same for $\Sigma_0$ when marginalising over $w_0$ and $\bar{\mu}_0$. Figure \ref{figure:withbao_2D} shows the $68.3\%$ forecast joint constraints on $\bar{\mu}_0-\Sigma_0$ in scenarios $1-3$ while marginalising over $w_0$ only. 

\begin{figure}[ht]
\begin{flushleft}
\includegraphics[width=0.8\linewidth]{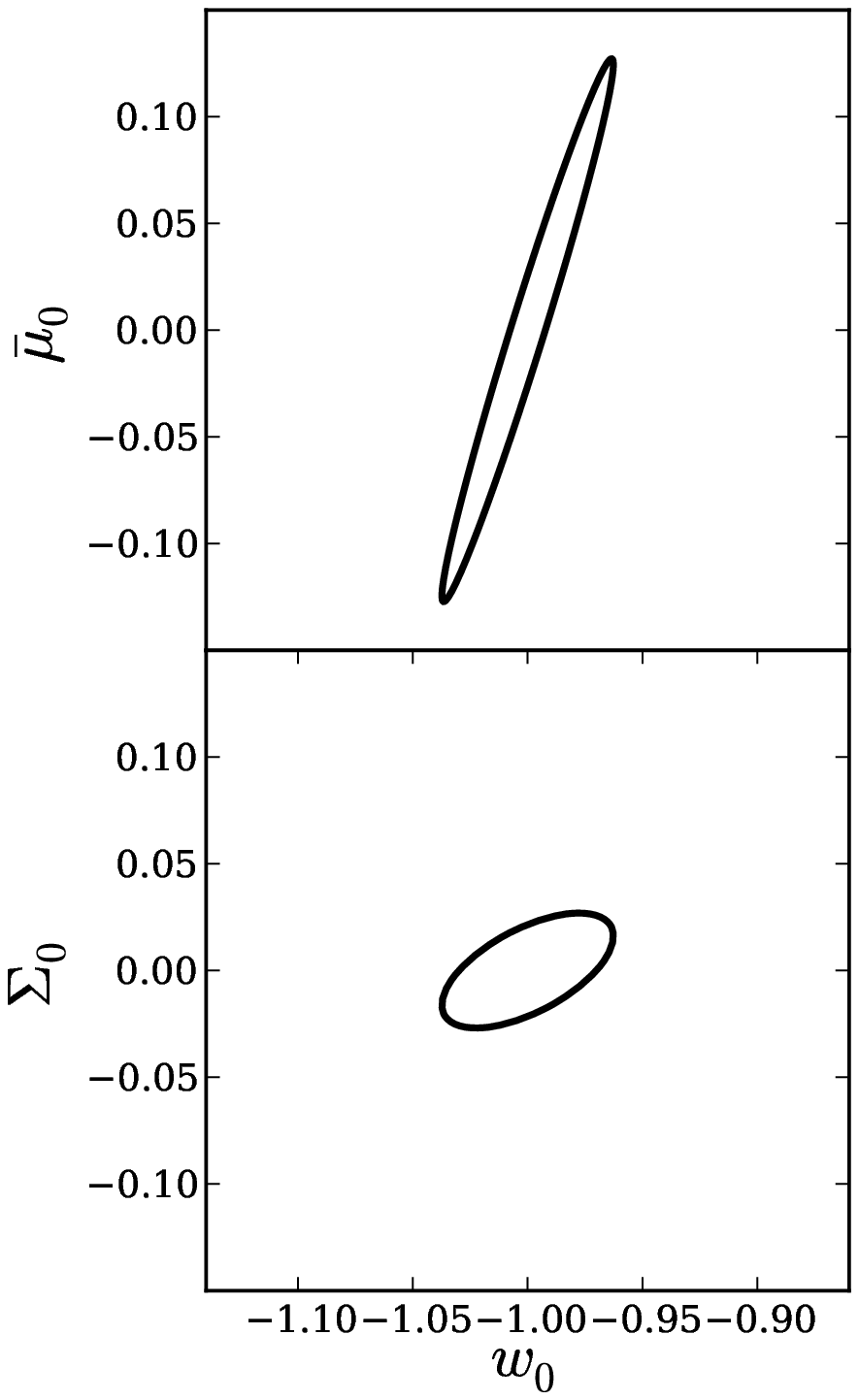}
\end{flushleft}
\caption{Forecast constraints at a $68.3 \%$ level from weak lensing, redshift-space distortions and BAO measurements, in the case where $\sigma_{w_0, BAO}=5\%$. The top panel displays these joint constraints in the $\bar{\mu}_0-w_0$ plane, marginalising over $\Sigma_0$, while the bottom panel does the same in the $\Sigma_0-w_0$ plane, marginalising over $\bar{\mu}_0$. In both cases, $w_a$ is fixed to $0$.}
\label{figure:marg2dpairs}
\end{figure}

We note from Figure \ref{figure:withbao_2D} that the degeneracy direction of the combined constraint in the $\bar{\mu}_0-\Sigma_0$ plane changes considerably between the three scenarios. $\bar{\mu}_0$ and $\Sigma_0$ are mildly negatively correlated in scenario $1$, whereas in scenario $2$ they are positively correlated, and in scenario $3$ even more so. This can be understood by considering the joint forecast constraints in the $\bar{\mu}_0-w_0$ and $\Sigma_0-w_0$ planes, marginalised in each case over the other non-$w_a$ parameter. These are displayed at a $68.3\%$ level in Figure \ref{figure:marg2dpairs} for scenario $3$. Both $\bar{\mu}_0$ and $\Sigma_0$ are shown therein to exhibit a positive correlation with $w_0$. This implies that $\bar{\mu}_0$ and $\Sigma_0$ are also positively correlated with each other, except in the case where $w_0$ is fixed or constrained so tightly that this effect is negated. As the constraint on $w_0$ is loosened, moving from scenario $1$ through scenario $2$ to scenario $3$, this positive correlation becomes more pronounced.

We notice also from Figure \ref{figure:withbao_2D} that the constraint on $\Sigma_0$ is relatively insensitive to the level of BAO constraint on $w_0$, whereas the constraint on $\bar{\mu}_0$ changes considerably between scenarios $1-3$.  This is consistent with Figure \ref{figure:marg2dpairs}, in which we see that the degeneracy direction in the $\bar{\mu}_0-w_0$ plane has a far greater positive slope than that in the $\Sigma_0-w_0$ plane. These degeneracy directions, and hence the relative sensitivity of $\bar{\mu}_0$ and $\Sigma_0$ constraints to $w_0$ constraints, can be understood by considering the expressions for $P_\kappa^{i,j}(\ell)$ (equation \ref{Pkappafinal}) and $\delta f\sigma_8(x)$ (equations \ref{deltaS_growth} and \ref{deltafsig8}).  Both $P_\kappa^{i,j}(\ell)$ and $\delta f\sigma_8(x)$ are given by integrals in time over a kernel and a source term.  In the case of $\delta f \sigma_8(x)$, the general relativistic kernel $G_f(x,\tilde{x})$ is significant back to $z\simeq 15$, whereas in the weak lensing case, the kernel is non-zero only as far back in redshift as the furthest source galaxies ($z=2$ in this case).  In the current model of $\beta(x)$, deviations from a $\Lambda$CDM expansion history are more significant at early times, whereas $\bar{\mu}(x)$ and $\Sigma(x)$ are both chosen to be significant only at late times (below $z\simeq 5$).  Therefore, the $\delta f\sigma_8(x)$ integration from $z\simeq 15$ favours sensitivity to the background expansion variable $w_0$ over $\bar{\mu}_0$, whereas the weak lensing integral, significant only from $z \simeq 2$, results in relatively greater sensitivity to $\Sigma_0$.  This results in the relative sensitivity of the $\bar{\mu}_0$ constraint to the $w_0$ constraint level, as seen in Figure \ref{figure:withbao_2D}.  Note that we have not accounted here for any uncertainty in galaxy bias models at high redshifts, which may have significant effects on the sensitivity of $f\sigma_8(x)$ to the background expansion at early times.

Finally, we note that there is clearly a direction in the $\bar{\mu}_0-\Sigma_0$ plane which is entirely insensitive to the change in $w_0$. This is in fact expected due to the nature of the contours displayed. Given the hypothetical 3D confidence region in the space of $\bar{\mu}_0$, $\Sigma_0$ and $w_0$, the marginalised constraint of scenario $3$ is equivalent to projecting this ellipsoid into the $\bar{\mu}_0-\Sigma_0$ plane. When we reduce the error in only the $w_0$ direction as in scenario $2$ -- that is, reducing the error in the direction orthogonal to the plane of projection -- the resulting projection will, by simple geometrical considerations, coincide with the first projection in two locations. The same argument can then be extended to the case of fixed $w_0$, which involves simply taking a slice of the 3D ellipsoid at the location of the $\bar{\mu}_0-\Sigma_0$ plane.

\subsubsection{Marginalising over $\{w_0, w_a\}$}
\label{subsubsection:wavaries}
\noindent
We now consider the case where we do not fix $w_a$ to zero. In this scenario, there is information present about $4$ parameters ($\mu_0, \Sigma_0, w_0, w_a$), so all Fisher matrices are $4 \times 4$. In addition to the previous derivative expressions, we now need derivatives with respect to $w_a$ of $P_\kappa^{i,j}(\ell)$ and $f\sigma_8(x)$. Once again, these are computed from equations \ref{Pkappafinal} and \ref{deltafsig8}, and listed in Appendix \ref{appendix:pderivs}. In this scenario, the BAO Fisher matrix is slightly more complicated, as the entire $2 \times 2$ block related to $w_0$ and $w_a$ is non-zero. 

In analogy to the above, we consider three scenarios:
\begin{enumerate}
\item{The scenario where $w_0$ and $w_a$ are fixed to their fiducial values.  Again, this for comparison, and is identical to the case considered in Section \ref{subsection:betazero_forecasts}.}
\item{The scenario where the BAO Fisher matrix represents the best-case expected constraints from a DETF4-type survey. In this scenario, the components of the BAO-only covariance matrix (the inverse of the Fisher matrix) are given by: $C_{w0,w0}=0.0010$, $C_{wa,w0}=-0.0038$, and $C_{wa,wa}=0.016$ \cite{Phil}. }
\item{The scenario where the BAO-only covariance matrix is obtained by multiplying the covariance matrix listed above in scenario $2$ by an overall factor of $(8.2)^2$.  This corresponds to the case where the projected $68.3\%$ error on $w_0$ from BAO is $5\%$ and all other elements of the covariance matrix are scaled up accordingly.}
\end{enumerate}

\begin{figure*}[t]
\begin{center}
\includegraphics[width=0.65\linewidth]{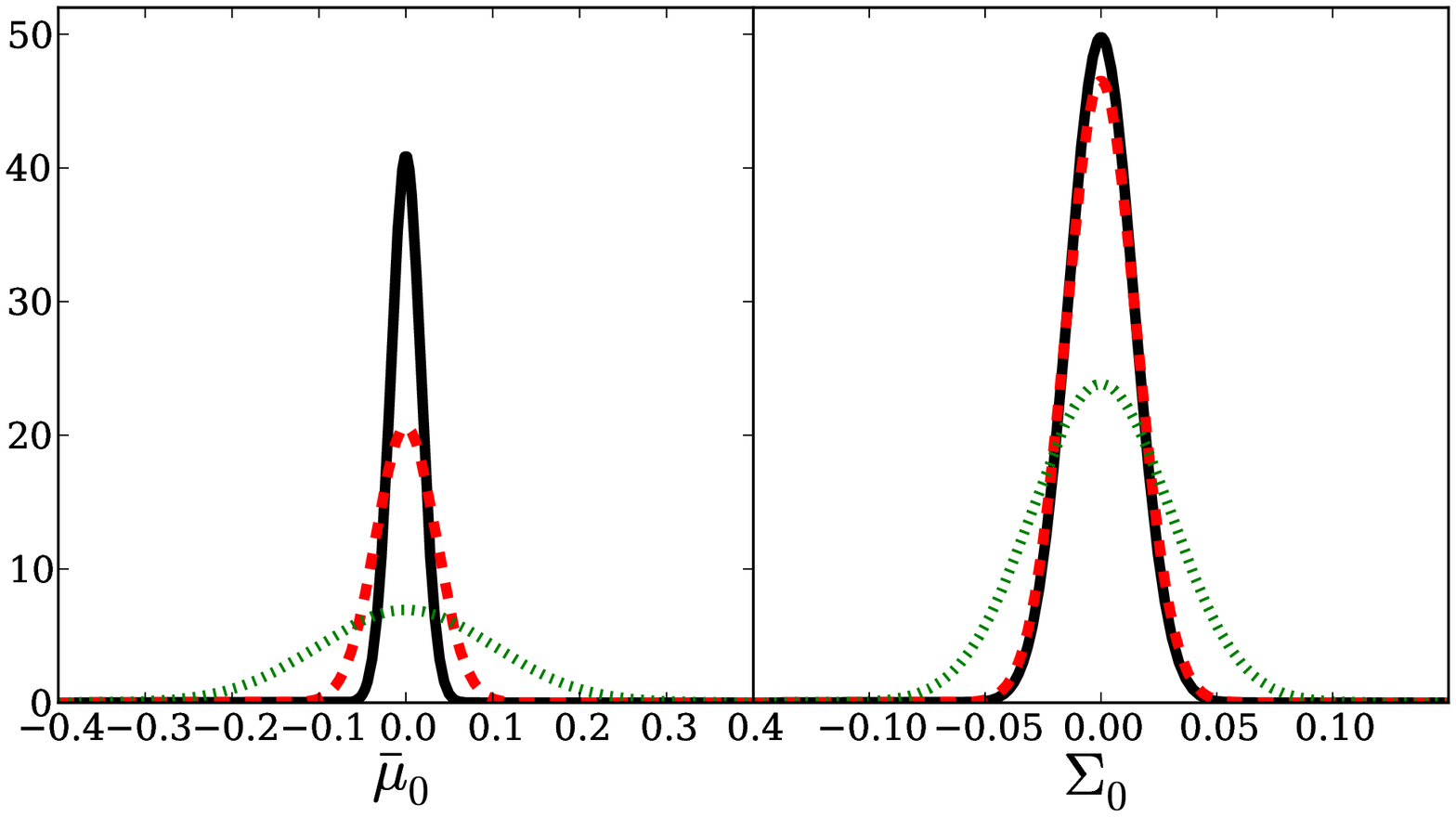}
\end{center}
\caption{Forecast constraints from weak lensing, redshift-space distortions, and BAO in the case where $w_0$ and $w_a$ have been marginalised over. The left-hand panel shows the confidence region for $\bar{\mu}_0$ when $\Sigma_0$ is marginalised over, while the right-hand panel shows the confidence region for $\Sigma_0$ with $\bar{\mu}_0$ marginalised over. Black, solid: $w_0$ and $w_a$ fixed; red, dashed: scenario $2$ described in Section \ref{subsubsection:wavaries}; green, dotted: scenario $3$ described in Section \ref{subsubsection:wavaries}.}
\label{figure:wamarg_1D}
\end{figure*}

\begin{figure}[ht]
\begin{center}
\includegraphics[width=\linewidth]{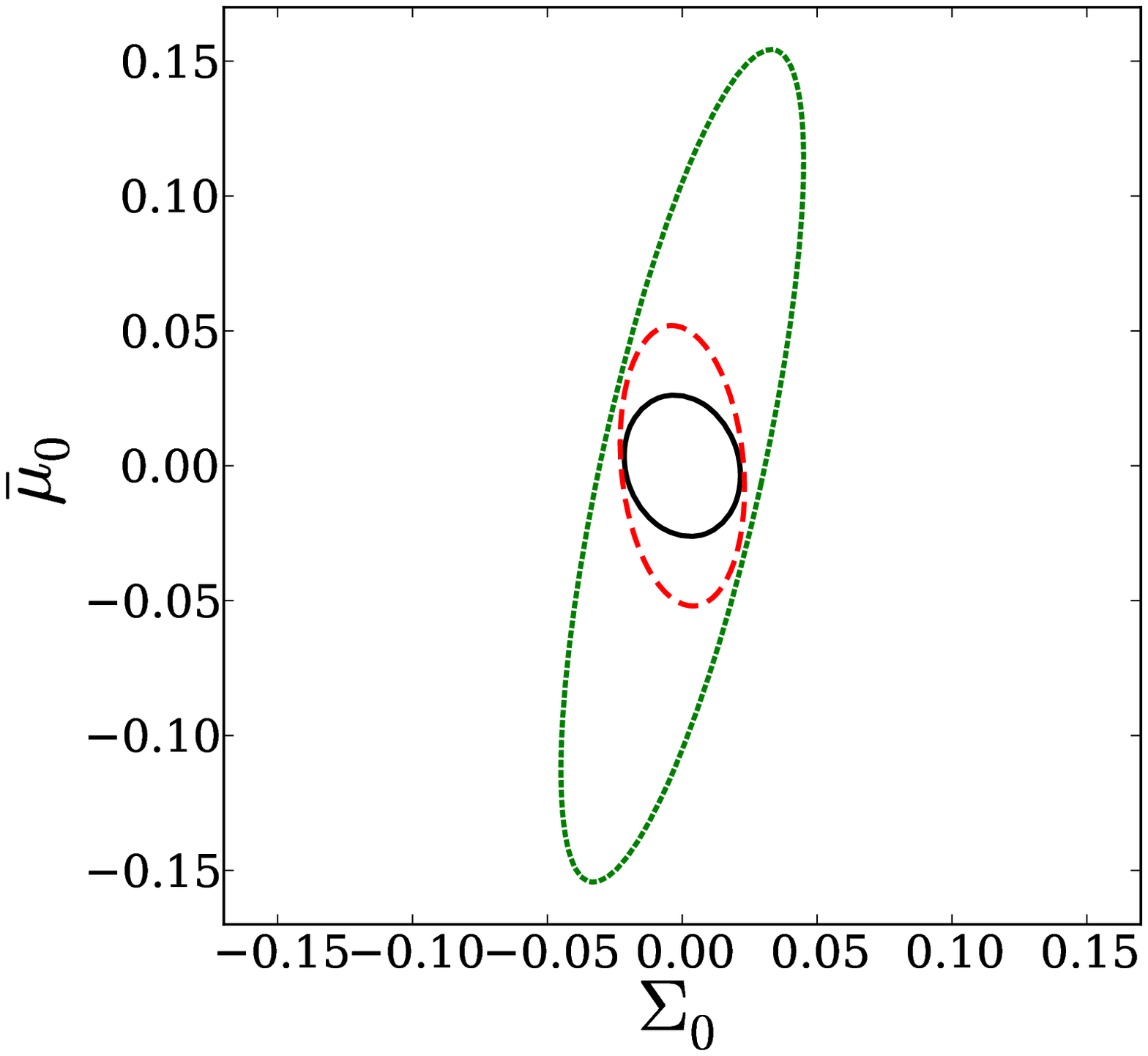}
\caption{Forecast $68.3\%$ confidence regions in the $\bar{\mu}_0-\Sigma_0$ plane, marginalising over $w_0$ and $w_a$. Black, solid: $w_0$ and $w_a$ fixed; red, dashed: scenario $2$ described in Section \ref{subsubsection:wavaries}; green, dotted: scenario $3$ described in Section \ref{subsubsection:wavaries}.}
\label{figure:wamarg_2D}
\end{center}
\end{figure}

The left-hand panel of Figure \ref{figure:wamarg_1D} presents the combined weak lensing, redshift-space distortion and BAO forecast constraints on $\bar{\mu}_0$ while marginalising over $w_0$, $w_a$ and $\Sigma_0$; the right-hand panel does the same for constraints on $\Sigma_0$ while marginalising over $w_0$, $w_a$ and $\bar{\mu}_0$. Figure \ref{figure:wamarg_2D}, meanwhile, presents the $68.3\%$ confidence regions in the $\bar{\mu}_0-\Sigma_0$ plane while marginalising over $w_0$ and $w_a$.

We see that the forecast constraint on $\Sigma_0$ is now slightly more sensitive to the level of BAO constraint on $w_0$ and $w_a$ than in the above case where $w_a$ is fixed. This is particularly noticeable in scenario $3$, in which the expansion history is the least well-constrained. Turning to $\bar{\mu}_0$, we see from Figure \ref{figure:wamarg_1D} that the forecast constraint remains sensitive to our knowledge of the expansion history in much the same way as in the $w_a$ fixed case. That is, the constraint in scenario $3$ is broadened considerably relative to that in scenario $2$, and both are slightly broader than in the above case where $w_a$ fixed.  Finally, examining the combined plot in Figure \ref{figure:wamarg_2D}, we see that the confidence regions therein are slightly larger than those in the corresponding Figure \ref{figure:withbao_2D}, where $w_a$ is fixed (other than for scenario $1$, which is the same in both figures by design). 

We surmise that allowing for a time-dependence in the equation of state of the effective dark energy component(via $\beta(x)$) loosens the expected constraints on $\bar{\mu}_0$ and $\Sigma_0$, but not catastrophically so. In fact, the level of constraint provided by BAO measurements on the expansion history of the universe appears to have a greater effect on forecast constraints in the $\bar{\mu}_0-\Sigma_0$ plane than does our assumption regarding the time-dependence of that expansion history.

\subsection{Scale-dependent $\mu(x,k)$ and $\gamma(x,k)$}
\label{subsection:scaledep_forecasts}
\noindent
Until this point, we have neglected any scale-dependence of $\mu(x,k)$ and $\gamma(x,k)$, focusing only on time-dependence. We now consider a scale-dependent ansatz.  

It has been shown that in the quasistatic regime and for local theories of gravity, $\mu(x,k)$ and $\gamma(x,k)$ can be expressed as a ratio of polynomials in $k$ with a specific form \cite{Silvestri2013}:
\begin{align}
\gamma(x,k)& \simeq \frac{p_1(x)+p_2(x)k^2}{1+p_3(x)k^2} \nonumber \\
\mu(x,k)& \simeq \frac{1+p_3(x)k^2}{p_4(x)+p_5(x)k^2}.
\label{silvestri}
\end{align}
This form has recently been considered in \cite{Hojjati2014}, in which a Principle Component Analysis was undertaken for a combined future data set including weak lensing and galaxy count measurements from the Large Synoptic Survey Telescope (LSST), as well as Planck measurements and upcoming supernova data.  Therein, the primary goal was to provide insight into future constraints on $\mu(x,k)$ and $\gamma(x,k)$ as well as on the overall scale-dependence. Here, we use the Fisher matrix formalism to find the level of forecast constraint on various parameter space directions from a DETF4-type survey. Our expression for $P^{i,j}_\kappa$ will then allow us to understand the related degeneracy structure of the parameter space.

Because we are interested only in broadly understanding the best-constrained directions of the parameter space, we fix $\beta(x)=0$ for simplicity. In GR+$\Lambda$CDM, $p_1(x)=p_4(x)=1$ and $p_2(x)=p_3(x)=p_5(x)=0$.  Therefore, in keeping with our linear response approach, we define:
\begin{align}
p_1(x)&=1+\delta p_1(x) \nonumber \\
p_2(x)&=\delta p_2(x) \nonumber \\
p_3(x)&=\delta p_3(x) \nonumber \\
p_4(x)&=1+\delta p_4(x) \nonumber \\
p_5(x)&=\delta p_5(x). 
\label{linrep_p}
\end{align}
Then, by dropping higher-order terms in $\delta p_i(x)$, we find expressions for $\mu(x,k)$ and $\gamma(x,k)$:
\begin{align}
\mu(x,k)& \simeq 1 + \delta p_3(x)k^2-\delta p_4(x)-\delta p_5(x) k^2\nonumber \\
\gamma(x,k)&\simeq1+\delta p_1(x)+\delta p_2(x) k^2- \delta p_3(x)k^2.
\label{silvestri_lr}
\end{align}
Our previous choices for $\delta \mu(x,k)$ and $\delta \gamma(x,k)$ have been consistent with equation \ref{silvestri_lr}; we have simply assumed that $\delta p_2(x)$, $\delta p_3(x)$, and $\delta p_5(x)$ have been sub-dominant, and let $-\delta p_4(x)=\delta \mu(x)$ and $\delta p_1(x)=\delta \gamma(x)$. 

It will be useful to have the equivalent expressions for  $\bar{\mu}(x,k)$ and $\Sigma(x,k)$.  We find them using equation \ref{param2}:
\begin{align}
\bar{\mu}(x,k)&\simeq -\delta p_1(x) - \delta p_4(x) \nonumber \\ &+k^2\left(-\delta p_2(x)+2\delta p_3(x)-\delta p_5(x)\right) \nonumber \\
\Sigma(x,k)&\simeq -\frac{1}{2} \delta p_1(x)-\delta p_4(x) \nonumber \\ &+ k^2\left(-\frac{1}{2}\delta p_2(x)+\frac{3}{2} p_3(x) - \delta p_5(x) \right).
\label{param2scaledep}
\end{align}

As usual, in order to forecast constraints using Fisher matrices, we must select an ansatz for the functions $\delta p_i(x)$. From equation \ref{silvestri}, we see that in fact, while $\delta p_1(x)$ and $\delta p_4(x)$ are dimensionless, $\delta p_2(x)$, $\delta p_3(x)$ and $\delta p_5(x)$ must have dimensions of length squared.  We choose a form similar to that of equation \ref{latetimemuSig} for both sets of functions, introducing a mass scale where necessary to account for their different dimensionalities:
\begin{align}
\delta p_{(1,4)}(x)&= p^{(1,4)}_0 \frac{ \Omega_\Lambda^{GR}(x)}{\Omega_\Lambda^{GR}(x=0)} \nonumber \\
\delta p_{(2,3,5)}(x)&= \frac{p^{(2,3,5)}_0 c^2}{(20H_0)^2} \frac{ \Omega_\Lambda^{GR}(x)}{\Omega_\Lambda^{GR}(x=0)}.
\label{silvestrip}
\end{align}
Thus, the parameter space is five-dimensional, with parameters $\{p^1_0, p^2_0, p^3_0, p^4_0, p^5_0\}$. The scale $c/20 H_0$ has been chosen as it provides sensible numerical eigenvalues.  We will see below that the numerical value of this scale is irrelevant to our results. 

To compute the Fisher matrices we require derivatives of $P^{i,j}_\kappa$ and $f\sigma_8$ with respect to the parameters $p_0^i$.  These are computed in a straightforward manner from equations \ref{Pkappafinal} and \ref{deltafsig8}, and are presented in Appendix \ref{appendix:pderivs}. As can be seen there, or as is obvious from equation \ref{silvestri}, some of these derivative expressions are dependent upon $k$.  In the weak lensing case, this is trivially dealt with, as the Limber approximation allows us to set $k=\frac{l}{\chi}$. However, the derivatives of $f\sigma_8$ with respect to $p^2_0$, $p^3_0$ and $p^5_0$ are truly $k$-dependent. 

In order to treat this case, we abandon the usual (general relativistic) notion that all scale-dependence of $f\sigma_8$ is factored out, and instead divide our forecast observations into five bins in $k$. As in \cite{Baker2014}, we let these bins have edges $k=[0.005, 0.02, 0.05, 0.08, 0.12, 0.15]$, stopping short of entering the regime where nonlinearities dominate. We select the error on the measurement in each $k$-bin based on the assumption that a DETF4-type survey, which covers a large fraction of the sky, will result in tighter measurements of large-scale modes than of small-scale modes.  To model this, we divide the total error budget of each redshift bin, given by \cite{Majerotto2012}, into the five $k$-bins listed above. The first redshift bin (between $k=0.005$ and $k=0.02$) receives $4\%$ of the error budget, the next two bins ($k=0.02-0.05$ and $k=0.05-0.08$) receive $12\%$ of the error budget each, and the final two bins ($k=0.08-0.12$ and $k=0.12-0.15$) receive $36\%$ of the error budget each.  

We can now construct the Fisher matrices for weak lensing and redshift-space distortions, recalling in the redshift-space distortions case to sum over bins of scale as well as of redshift. We invert the total combined Fisher matrix to obtain the covariance matrix, and diagonalise to obtain five eigenvectors with corresponding eigenvalues.

Before examining and interpreting these eigenvalues and eigenvectors, we pause to consider more fully the implications of the difference in dimensionality of $\delta p_1(x)$ and $\delta p_4(x)$ vs $\delta p_2(x)$, $\delta p_3(x)$ and $\delta p_5(x)$.  In equation \ref{silvestrip}, $p_2^0$, $p_3^0$ and $p_5^0$ have essentially been scaled by $\frac{c^2}{(20H_0)^2}$. This scaling factor is arbitrary, and using a different scaling factor would alter the the numerical value of the eigenvalues of the covariance matrix.  Therefore, although all eigenvalues of the covariance matrix are dimensionless, any forecast constraint on $p_2^0$, $p_3^0$ or $p_5^0$, or on any combination thereof, must be multiplied by $\frac{c^2}{(20 H_0)^2}$ (or the appropriate scaling factor). This provides a physically meaningful value with dimensions length squared, which represents a constraint on the combination $p_{(2,3,5)}^0 \times \frac{c^2}{(20 H_0)^2}$.  Without introducing a prior, it is impossible to meaningfully compare numerical constraints on parameter combinations within two different parameter sub-spaces: that of $p_1^0$ and $p_4^0$ and that of $p_2^0$, $p_3^0$, and $p_5^0$. 

First, consider those eigenvectors in the subspace of $p_1^0$ and $p_4^0$.  We list them here, along with the square root of the associated eigenvalue $\sqrt{\sigma^2}$.  This acts as a measure of how well the parameter space is constrained in the direction of the eigenvector.  We have:
\begin{align}
\vec{\alpha}_1&=0.8\hat{p}_1-0.004\hat{p}_2+0.001\hat{p}_3+0.6 \hat{p}_4+0.002 \hat{p}_5 \nonumber \\
& \simeq 0.8\hat{p}_1+0.6 \hat{p}_4
\label{alpha1}
\end{align}
with $\sqrt{\sigma^2} \simeq 0.01$, and 
\begin{align}
\vec{\alpha}_2&=-0.6\hat{p}_1+0.0003\hat{p}_2+0.0002\hat{p}_3+0.8\hat{p}_4-0.0006\hat{p}_5 \nonumber \\
&\simeq -0.8\hat{p}_1+0.6\hat{p}_4
\end{align}
with $\sqrt{\sigma^2} \simeq 0.02$. Clearly, due to the numerical nature of the Fisher matrix calculation, some sub-dominant contributions in the directions of $\hat{p}_2$, $\hat{p}_3$ and $\hat{p}_5$ persist; we ignore these.  We see here that $\vec{\alpha}_1$, the better constrained of the two parameter space directions, is essentially a weighted sum of the directions $\hat{p}_1$ and $\hat{p}_4$. $\vec{\alpha}_2$ is the vector orthogonal to $\vec{\alpha}_1$, and consists of a weighted difference.  

In order to interpret the level of constraint placed on the eigenvector directions, we consider equation \ref{param2scaledep}.  Recall that because we have chosen an ansatz for $\delta p_i(x)$ which is non-negligible at late times only (equation \ref{silvestrip}), $\Sigma(x,k)$ is the dominant non-GR+$\Lambda$CDM contribution to $P_\kappa^{i,j}(\ell)$, while only $\bar{\mu}(x,k)$ contributes to $f\sigma_8$. In examining equation \ref{param2scaledep}, it is clear why a sum of directions $\hat{p}_1$ and $\hat{p}_4$ is forecast to be better constrained than a difference: both $\Sigma(x,k)$ and $\bar{\mu}(x,k)$ are directly sensitive to weighted sums of $\delta p_1(x)$ and $\delta p_4(x)$.  

We now consider eigenvectors in the subspace of $p_2^0$, $p_3^0$ and $p_5^0$.  These are given by:
\begin{align}
\vec{\alpha}_3&=0.0005\hat{p}_1+0.3\hat{p}_2-0.8\hat{p}_3+0.0009\hat{p}_4+0.5\hat{p}_5 \nonumber \\
&\simeq 0.3\hat{p}_2-0.8\hat{p}_3+0.5\hat{p}_5 
\end{align}
with $\sqrt{\sigma^2} \simeq 10^{-5}$,
\begin{align}
\vec{\alpha}_4&=-0.004 \hat{p}_1-0.8\hat{p}_2+0.1\hat{p}_3-0.002 \hat{p}_4+0.6\hat{p}_5 \nonumber \\
& \simeq -0.8\hat{p}_2+0.1\hat{p}_3+0.6\hat{p}_5
\end{align}
with $\sqrt{\sigma^2} \simeq 4 \times 10^{-4}$, and
\begin{align}
\vec{\alpha}_5&=-1\times 10^{-6}\hat{p}_1+0.6\hat{p}_2+0.6\hat{p}_3+5\times10^{-5}\hat{p}_4+0.6\hat{p}_5 \nonumber \\
&\simeq 0.6\hat{p}_2+0.6\hat{p}_3+0.6\hat{p}_5
\end{align}
with $\sqrt{\sigma^2} \simeq 2$.  Recall that it is not meaningful to directly compare these $\sqrt{\sigma^2}$ values with those for eigenvectors $\vec{\alpha}_1$ and $\vec{\alpha}_2$. 

In order to understand the relative constraints on the eigenvectors in this subspace, we first find the relationships between parameter values along the eigenvector directions. We then use these relationships in equation \ref{param2scaledep} to find simpler expressions for $\Sigma(x,k)$ and $\bar{\mu}(x,k)$ :
\begin{itemize}
\item{In the direction $\vec{\alpha}_3$, $p_2^0 \simeq -\frac{1}{3}p_3^0$ and $p_5^0 \simeq -\frac{2}{3}p^0_3$.  Therefore, $\Sigma(x,k) \simeq \frac{7}{3}\frac{p^{3}_0 c^2}{(20H_0)^2} \frac{ \Omega_\Lambda^{GR}(x)}{\Omega_\Lambda^{GR}(x=0)}k^2$, and $\bar{\mu}(x,k) \simeq 3\frac{p^{3}_0 c^2}{(20 H_0)^2} \frac{ \Omega_\Lambda^{GR}(x)}{\Omega_\Lambda^{GR}(x=0)}k^2$.}
\item{In the direction  $\vec{\alpha}_4$, $p_2^0 \simeq -8p_3^0$ and $p_5^0 \simeq 6p^0_3$. $\Sigma(x,k) \simeq -\frac{1}{2}\frac{p^{3}_0 c^2}{(20H_0)^2} \frac{ \Omega_\Lambda^{GR}(x)}{\Omega_\Lambda^{GR}(x=0)}k^2$, and $\bar{\mu}(x,k) \simeq 4\frac{p^{3}_0 c^2}{H_0^2} \frac{ \Omega_\Lambda^{GR}(x)}{\Omega_\Lambda^{GR}(x=0)}k^2$.}
\item{In the direction $\vec{\alpha}_5$, $p_2^0 \simeq p_3^0$ and $p_5^0 \simeq p^0_3$.  $\Sigma(x,k) \simeq 0$, and $\bar{\mu}(x,k) \simeq 0$.}
\end{itemize}
With this manipulation, it can be seen why the direction $\vec{\alpha}_3$ is better constrained than $\vec{\alpha}_4$. The expressions for $\Sigma(x,k)$ demonstrate that a small change in parameter values along direction $\vec{\alpha}_3$ has a numerically larger effect on the value of $\Sigma(x,k)$ than does a small change along direction $\vec{\alpha}_4$, by a factor of $\frac{14}{3}$.  Although $\bar{\mu}(x,k)$ is affected slightly more by a change along the $\vec{\alpha}_4$ direction than along the $\vec{\alpha}_3$ direction (by a factor of $\frac{4}{3}$), this effect is clearly subdominant. It is also plain to see why the direction $\vec{\alpha}_5$ is by far the worst constrained of this set. As long as $p_2^0=p_3^0=p_5^0$, both $\Sigma(x,k)$ and $\bar{\mu}(x,k)$ are entirely insensitive to the value taken by these parameters.  

It is tempting to attempt a comparison of the above results directly with those of \cite{Hojjati2014}. However, this would be misleading, as the Principle Component Analysis employed in that work allows the functions $p_i(x)$ to take any time-dependence, whereas we restrict the time-dependence to that given in equation \ref{silvestrip}.  Additionally, the authors of \cite{Hojjati2014} choose to impose a prior upon the variance of each function $p_i(x)$, enabling them to meaningfully discuss constraints on directions which combine the two parameter subspaces which we have considered. Nevertheless, there is some small comparison we can draw: in \cite{Hojjati2014} it was found that the functions $p_i(x)$ can not be individually constrained, and we similarly find no evidence of any well-constrained direction corresponding to a single $p_0^i$.

\section{Conclusions}
\label{section:discussion}
\noindent
The goal of this work has been to understand how weak lensing measurements are affected by the individual physical effects of alternative theories of gravity. In this spirit, we have chosen to prioritise clarity.  We have therefore restricted ourselves to considering gravitational parameters, and have not included uncertainty in galaxy bias or in the standard cosmological parameter values at this time.

We have constructed an expression for the power spectrum of the weak lensing observable convergence, as given in equation \ref{Pkappafinal}.  By considering only small deviations from GR+$\Lambda$CDM, we have derived an expression which separates into an integral over two terms: a general relativistic kernel, and a source term which encompasses all deviations from general relativity.  This source term is composed of additive terms which are themselves each representative of a different effect due to modifying general relativity.  This neat separation, first of the full expression into kernel and source and then of the source expression into physical effects, allows degeneracies between gravitational parameters to be physically interpreted.  We note also that our expression is reliable in the $\Lambda$CDM-like case of $\beta(x)=0$, whereas complimentary works employing a fluid model have found this limit difficult to constrain \cite{Battye:2014xna}.

With equation \ref{Pkappafinal} in hand, we first investigated degeneracies in the simplified case of a $\Lambda$CDM-like expansion history, showing how the degeneracy direction of weak lensing in the $\bar{\mu}_0-\Sigma_0$ plane relies on the time-dependent ansatz for $\bar{\mu}(x)$.  However, we also demonstrated that even for an extreme ansatz, weak lensing and redshift-space distortion measurements remain non-degenerate in this plane, and hence are a viable combination of observables to offer joint constrains on $\bar{\mu}_0$ and $\Sigma_0$.  

We then moved on to exploit the potential of our expression as a valuable tool in conducting and interpreting Fisher forecasts of weak lensing observations. We found that the linearity of our expression in the gravitational parameters meant that it was technically simple to compute the required Fisher matrices. Perhaps more importantly, the simple form of our expression was also found to provide physical interpretation of the degeneracies which presented themselves in the forecast of multi-dimensional constraints.

We first demonstrated this use of our expression by allowing the effective equation of state of the dark energy component to take a CPL form, given in Section \ref{subsection:bao_forecasts}.  We found that in the case where $w_a=0$ and a phenomenological ansatz for $\bar{\mu}(x)$ and $\Sigma(x)$ is assumed, forecast constraints on $\bar{\mu}_0$ from weak lensing and redshift-space distortions are highly sensitive to the level of constraint on $w_0$ from BAO. Forecast constraints on $\Sigma_0$, on the other hand, were found to be nearly unaffected. The separation of our expression into source and kernel made evident the stark difference between the redshift dependences of the kernels of $P_\kappa^{i,j}$ and $f\sigma_8$. This provided a clear explanation for the higher sensitivity of redshift-space distortions to early time effects, and hence to changes in $w_0$. Relaxing the requirement that $w_a=0$, we then saw that constraints in the $\bar{\mu}_0-\Sigma_0$ plane were only moderately affected by marginalising over $w_a$.  This offers the exciting prospect that the analysis of data from a DETF4-type survey may be able to allow for a time-dependent background expansion without sacrificing much constraining power in the $\bar{\mu}_0-\Sigma_0$ plane.  

Finally, we used our expression for $P_\kappa^{i,j}$, in combination with that for $f\sigma_8$, to understand the best-constrained directions in the space of parameters of the scale-dependent ansatz given in \cite{Silvestri2013}. By varying the parameter values along the eigenvectors of the Fisher matrix within the source terms of $P_\kappa^{i,j}$ and of $f\sigma_8$, we could understand analytically why different directions in the parameter space were forecast to be well- or poorly-constrained. Using again a linear response approach, we were able to explain why a weighted sum of the parameters not associated with scale-dependence is forecast to be well-constrained, and why a direct sum of the parameters associated with scale-dependence is totally unconstrained. This work expands upon the work of \cite{Hojjati2014}, both by forecasting for a different survey, but also by providing clear reasons for the forecast constraints found, something that is non-trivial to do using a Principle Component Analysis method.  

The methods we have developed to better understand the effect of altering the theory of gravity could in principle be extended to understanding degeneracies in other scenarios of observational cosmology.  We hope that we have provided here the groundwork for such developments.  As previously noted, in the interest of clarity we have not included uncertainty on the galaxy bias or on standard cosmological parameters in our forecasting. Ideally, these would be included to provide more accurate forecasting and more general conclusions.  However, in seeking to include these additional parameters, the clarity of equation \ref{Pkappafinal} is compromised, and as such we anticipate that future work in this direction will necessarily involve the sacrifice of the descriptive power shown here. In such future work, forecasting methods involving sampling the posterior of the joint probability distribution of the relevant parameters are likely to be more appropriate.

\section*{Acknowledgements}

We would like to thank Phil Bull, Lance Miller and Catherine Heymans for helpful discussions. We also thank the authors of the publicly available code CAMB, which was used in this work. CDL is supported by the Rhodes Trust. TB is supported by All Souls College, University of Oxford.  PGF acknowledges support from  STFC, BIPAC, a Higgs visiting fellowship and the Oxford Martin School.

\appendix
\section{Derivation of $\Hu^2\Omega_M|_{MG} = \Hu^2\Omega_M|_{GR}$}
\label{appendix:H2Om}

Here we derive a) the formula for $\delta\Hu(x)$ used in equation \ref{delH}, and b) the relation $\Hu^2\Omega_M|_{MG} = \Hu^2\Omega_M|_{GR}$ used in Section \ref{subsection:convergenceMG}.

First we write the Friedmann equation as:
\begin{align}
\Hu^2(x)&=\frac{8\pi G}{3}a^2\left[\rho_M(a)+\rho_D(a)\right]\nonumber\\ 
&=\frac{8\pi G}{3}\left[\rho_{M0}e^{-x}+\rho_{D 0}e^{-\int dx' \,(1+3 w_{D}(x'))}\right]\nonumber\\
&=H_0^2\Omega_{M0}e^{-x}\left[1+Re^{3x}e^{-3\int_0^x dx' \,\beta(x')}\right]
\end{align}
where in the second line we have changed the independent variable to $x=\ln a$, and used the energy density evolution  for a fluid with a general equation of state $\omega_D$. In the third line we have defined $R=\Omega_{\Lambda 0}/\Omega_{M0}$, identifying the present fractional energy density in the dark fluid ($\Omega_{D0}$) with that of the apparent cosmological constant today ($\Omega_{\Lambda0}$). We have also made use of $w_D(x)=-1+\beta(x)$, as introduced in equation \ref{defbeta}. 

We define $u(x)=\int_0^x \beta(x') \,dx'$ as in equation \ref{u}, and assume as in Section \ref{subsection:convergenceMG} that for a viable cosmology $|\beta(x)| \ll 1$ and $|u(x)|\ll 1$. Then, expanding the exponential and taking a square root leads to:
\begin{align}
\Hu(x)&=H_0\sqrt{\Omega_{M0}}e^{-\frac{x}{2}}\left[1+Re^{3x}(1-3u(x))\right]^{\frac{1}{2}}\nonumber\\
&\approx H_0\sqrt{\Omega_{M0}}e^{-\frac{x}{2}}\left[1+Re^{3x}\right]^{\frac{1}{2}}\left[1-\frac{3}{2}\left(\frac{u(x)Re^{3x}}{1+Re^{3x}}\right)\right]\nonumber\\
&=\Hu_{GR}(x)\left[1-\frac{3}{2}\left(\frac{u(x)Re^{3x}}{1+Re^{3x}}\right)\right]
\end{align}
where in the second line we have used a Taylor expansion to linear order. Finally, we use the (easily derived) result that in $\Lambda$CDM, and assuming negligible radiation,
\begin{align}
\left(1-\Omega_M^{GR}\right)&=\frac{Re^{3x}}{1+Re^{3x}}
\end{align}
to obtain:
\begin{align}
\label{appeq1}
\delta \Hu(x)&=\Hu(x)-\Hu_{GR}(x)=-\frac{3}{2}\Hu_{GR}(x)u(x)\left(1-\Omega_M^{GR}\right).
\end{align}
With this result in hand, it is simple to show that the combination $\Hu^2\Omega_M$ does not change under the perturbations about the GR+$\Lambda$CDM model. Expanding to first order about the fiducial model:
\begin{align}
\Hu^2\Omega_M&=\left(\Hu_{GR}+\delta\Hu\right)^2\left(\Omega^{GR}_M+\delta\Omega_M\right)\\
&\approx\Hu_{GR}^2\Omega_M^{GR}\left(1+2\frac{\delta\Hu}{\Hu_{GR}}+\frac{\delta\Omega_M}{\Omega_M^{GR}}\right)+{\cal O}(\delta\Hu^2).
\label{appeq2}
\end{align}
We call upon a result derived in Appendix A of \cite{Baker2014} (for brevity's sake we will not repeat the derivation here):
\begin{align}
\label{appeq3}
\delta\Omega_M&=3u(x)\Omega_M^{GR} \left[1-\Omega_M^{GR}\right].
\end{align}
Substituting equations (\ref{appeq1}) and (\ref{appeq3}) into equation (\ref{appeq2}) one finds $\Hu^2\Omega_M=\Hu_{GR}^2\Omega_M^{GR}$, as stated in the text.

\section{Converting between $\{\delta\mu(x,k), \delta \gamma(x,k)\}$ and $\{\bar{\mu}(x,k), \Sigma(x,k)\}$}
\label{appendix:convertfuncs}
\noindent
We derive here the relationship, given in equation \ref{param2}, between two sets of functions which parameterise deviations from GR+$\Lambda$CDM in the quasistatic limit: $\{\delta\mu(x,k), \delta \gamma(x,k)\}$ and $\{\bar{\mu}(x,k), \Sigma(x,k)\}$.  In doing so, we will refer heavily to equations 4 and 5 of \cite{Simpson2012}. For reference, we reproduce them here, changing to our time variable $x=\ln(a)$ and making some minor notational alterations in order to maintain our conventions:
\begin{align}
\Psi(x,k)&=[1+\bar{\mu}(x,k)]\Psi^S_{GR}(x,k) \nonumber \\
\Psi(x,k)+\Phi(x,k)&=[1+\Sigma(x,k)](\Psi^S_{GR}(x,k)+\Phi^S_{GR}(x,k)).
\label{simpsondefs}
\end{align}
We have labeled the general relativistic potentials with an $S$ due to the fact that they are slightly different from what we call $\Psi_{GR}(x,k)$ and $\Phi_{GR}(x,k)$.  $\Psi^S_{GR}(x,k)$ and $\Phi^S_{GR}(x,k)$ follow a general relativistic Poisson equation and slip relation, but they may generally still have a different value than $\Psi_{GR}(x,k)$ and $\Phi_{GR}(x,k)$, due to any difference in the history of the growth of overdensities.

Now, from equation \ref{poissoneqn} above, we see that 
\begin{equation}
\Phi(x,k)=(1+\delta \mu(x,k) ) (1+ \delta_\Delta(x,k))\Phi_{GR}(x,k).
\end{equation}
In fact, $\Phi^S_{GR}(x,k)=(1+ \delta_\Delta(x,k))\Phi_{GR}(x,k)$, so that we have:
\begin{equation}
\Phi(x,k)=(1+\delta \mu(x,k) )\Phi^S_{GR}(x,k).
\label{phisimp}
\end{equation} 
We also know from equation \ref{defgammu} that in the quasistatic regime, $\Phi(x,k)=(1+\delta \gamma(x,k))\Psi(x,k)$.  Therefore:
\begin{align}
\Psi(x,k)&=\frac{1+\delta \mu(x,k)}{1+\delta \gamma(x,k)}\Phi^S_{GR}(x,k) \nonumber \\
&=\frac{1+\delta \mu(x,k)}{1+\delta \gamma(x,k)}\Psi^S_{GR}(x,k) \nonumber \\
&\simeq(1+\delta \mu(x,k)-\delta \gamma(x,k))\Psi^S_{GR}(x,k)
\label{psisimp}
\end{align}
where the second line comes from the fact that $\Phi^S_{GR}(x,k)=\Psi^S_{GR}(x,k)$.  Referring to equation \ref{simpsondefs}, we see that
\begin{equation}
\bar{\mu}(x,k)=\delta \mu(x,k)-\delta \gamma(x,k),
\label{defbarmu}
\end{equation}
as in equation \ref{param2}.  

From here, we can use equations \ref{phisimp} and \ref{psisimp} to write (suppressing time- and scale-dependence for brevity):
\begin{align}
\Psi+\Phi&=\frac{1+\delta\mu}{1+\delta \gamma}\Psi^S_{GR}+(1+\delta \mu)\Phi^S_{GR} \nonumber \\
&=\frac{1+\delta\mu}{1+\delta \gamma}\Phi^S_{GR}+(1+\delta \mu)\Phi^S_{GR} \nonumber \\
&=\frac{1}{2}\left[\frac{1+\delta\mu}{1+\delta \gamma}+(1+\delta \mu)\right](\Phi^S_{GR}+\Psi^S_{GR}) \nonumber \\
& \simeq \left(1+\delta\mu-\frac{1}{2}\delta \gamma\right)(\Phi^S_{GR}+\Psi^S_{GR}) 
\end{align}
and by comparison with equation \ref{simpsondefs}, we have that 
\begin{equation}
\Sigma(x,k)=\delta\mu(x,k)-\frac{1}{2}\delta \gamma(x,k)
\end{equation}
as in equation \ref{param2}.

\section{Derivatives of $P^{i,j}_\kappa$ and $f\sigma_8$}
\label{appendix:pderivs}
\noindent
We present here the derivatives of $P_\kappa^{i,j}$ and $f\sigma_8$ which are required for forecasting in Section \ref{section:constraints}. These are determined in a straightforward manner by differentiating equations \ref{Pkappafinal} and \ref{deltafsig8}. For brevity, we will use $\mathcal{K}(x,\ell)$ as defined in equation \ref{K} to denote the general relativistic kernel in the definition of $P_\kappa^{i,j}(\ell)$, and $G_f(x,\tilde{x})$ as defined in equation 34 of \cite{Baker2014} to denote the general relativistic kernel in the case of redshift-space distortions. 

First, the derivatives with respect to $\bar{\mu}_0$ and $\Sigma_0$ are given as:
\begin{widetext}
\begin{align}
\frac{\partial P^{i,j}_\kappa(\ell)}{\partial \bar{\mu}_0}&=\int_{-\infty}^{0}dx\,\mathcal{K}(x,\ell) \left[3 \int_{-\infty}^{x} d \bar{x} \,
  \Omega_{M}^{GR}(\bar{x}) I(x,\bar{x}) \frac{\Omega_{\Lambda}^{GR}(\bar{x})}{\Omega_{\Lambda}^{GR}(\bar{x}=0)} \right] \nonumber \\
\frac{\partial
  P^{i,j}_\kappa(\ell)}{\partial \Sigma_0}&=
2\int_{-\infty}^{0}dx\, \mathcal{K}(x,\ell)\frac{\Omega^{GR}_{\Lambda}(x)}{\Omega^{GR}_{\Lambda}(x=0)} \nonumber \\
\frac{\partial f\sigma_8(x)}{\partial \bar{\mu}_0} &=\int_{-\infty}^{x} G_f(x,\tilde{x})\frac{\Omega^{GR}_{\Lambda}(\tilde{x})}{\Omega^{GR}_{\Lambda}(\tilde{x}=0)} d\tilde{x} \nonumber \\
\frac{\partial f\sigma_8(x)}{\partial \Sigma_0} &= 0.
\label{diffmuSig}
\end{align}
Next, derivatives with respect to $w_0$ and $w_a$:
\begin{align}
\frac{\partial P^{i,j}_\kappa(\ell)}{\partial w_0} &=\int_{-\infty}^0 dx \,\mathcal{K}(x,\ell)\Bigg\{\frac{3}{2}\int_{-\infty}^{x} d\bar{x}\,I(x,\bar{x})(1-\Omega_M^{GR}(\bar{x}))\left[3\Omega_M^{GR}(\bar{x})(1+f_{GR}(\bar{x}))\bar{x}+f_{GR}(\bar{x})\right]+ \frac{3x}{2}(1-\Omega^{GR}_M(x)) \nonumber \\ 
&+ \left(\frac{\partial \ln G_j(x)}{\partial \ln \chi} + \frac{\partial \ln G_i(x)}{\partial \ln \chi}  - \frac{\partial \ln (P^{GR}_\delta (x=0, k)/k^4)}{\partial \ln k}\right) \Bigg|_{\chi_{GR}} \frac{3}{2\chi_{GR}(x)} \int_{\infty}^{x}d\bar{x} \frac{\bar{x}}{\mathcal{H}_{GR}(\bar{x})}(1-\Omega^{GR}_{M}(\bar{x})) \Bigg\} \nonumber \\
\frac{\partial P^{i,j}_\kappa(\ell)}{\partial w_a} &= \int_{-\infty}^0 dx \,\mathcal{K}(x,\ell) \Bigg( \frac{3}{2}[x-(e^{x}-1)](1-\Omega^{GR}_M(x)) \nonumber \\ &+\frac{3}{2}\int_{-\infty}^{x} d\bar{x}\, I(x,\bar{x})(1-\Omega_M^{GR}(\bar{x}))\left\{3\Omega_M^{GR}(\bar{x})(1+f_{GR}(\bar{x}))[\bar{x}-(e^{\bar{x}}-1)]+f_{GR}(\bar{x})(1-e^{\bar{x}})\right\} \nonumber \\ &+ \left(\frac{\partial \ln G_j(x)}{\partial \ln \chi} + \frac{\partial \ln G_i(x)}{\partial \ln \chi}  - \frac{\partial \ln (P^{GR}_\delta (x=0, k)/k^4)}{\partial \ln k}\right) \Bigg|_{\chi_{GR}} \frac{3}{2} \int_{\infty}^{x}d\bar{x} \frac{\bar{x}-(e^{\bar{x}}-1)}{\mathcal{H}_{GR}(\bar{x})}(1-\Omega^{GR}_{M}(\bar{x})) \Bigg) \\
\frac{\partial f\sigma_8(x)}{\partial w_0}  &= \int_{-\infty}^{x} G_f(x,x')\left[\frac{(1-\Omega_{M}^{GR}(x'))}{\Omega_{M}^{GR}(x')}3\Omega_{M}^{GR}(x')(1+f_{GR}(x'))x'+f_{GR}(x')\right] dx' \nonumber \\
\frac{\partial f\sigma_8(x)}{\partial w_a}  &= \int_{-\infty}^{x} G_f(x,x')\left\{\frac{(1-\Omega_{M}^{GR}(x'))}{\Omega_{M}^{GR}(x')}3\Omega_{M}^{GR}(x')(1+f_{GR}(x'))[x'-(e^{x'}-1)]+f_{GR}(x')(1-e^{x'})\right\} dx'.
\label{w0waderivs}
\end{align}
Finally, derivatives with respect to $\{ p_0^i \}$:
\begin{align}
\frac{\partial P^{i,j}_\kappa(\ell)}{\partial p^1_0} &= -\int_{-\infty}^0 dx \, \mathcal{K}(x,\ell)\left[ \frac{\Omega^{GR}_\Lambda(x)}{\Omega^{GR}_\Lambda(x=0)}+3 \int_{-\infty}^x dx' \, I(x,x') \Omega^{GR}_M(x')\frac{\Omega^{GR}_\Lambda(x')}{\Omega^{GR}_\Lambda(x=0)} \right] \nonumber \\
\frac{\partial P^{i,j}_\kappa(\ell)}{\partial p^2_0} &= -\int_{-\infty}^0 dx \, \mathcal{K}(x,\ell)\left[ \frac{\Omega^{GR}_\Lambda(x)}{\Omega^{GR}_\Lambda(x=0)}\frac{\ell^2}{\chi_{GR}^2(x)}+3 \int_{-\infty}^x dx' \, I(x,x') \Omega^{GR}_M(x')\frac{\Omega^{GR}_\Lambda(x')}{\Omega^{GR}_\Lambda(x=0)}\frac{\ell^2}{\chi_{GR}^2(x')} \right] \nonumber \\
\frac{\partial P^{i,j}_\kappa(\ell)}{\partial p^3_0} &= \int_{-\infty}^0 dx \, \mathcal{K}(x,\ell)\left[ 3\frac{\Omega^{GR}_\Lambda(x)}{\Omega^{GR}_\Lambda(x=0)}\frac{\ell^2}{\chi_{GR}^2(x)}+6 \int_{-\infty}^x dx' \, I(x,x') \Omega^{GR}_M(x')\frac{\Omega^{GR}_\Lambda(x')}{\Omega^{GR}_\Lambda(x=0)}\frac{\ell^2}{\chi_{GR}^2(x')} \right] \nonumber \\
\frac{\partial P^{i,j}_\kappa(\ell)}{\partial p^4_0} &= -\int_{-\infty}^0 dx  \, \mathcal{K}(x,\ell)\left[ 2\frac{\Omega^{GR}_\Lambda(x)}{\Omega^{GR}_\Lambda(x=0)}+3 \int_{-\infty}^x dx' \, I(x,x') \Omega^{GR}_M(x')\frac{\Omega^{GR}_\Lambda(x')}{\Omega^{GR}_\Lambda(x=0)} \right] \nonumber \\
\frac{\partial  P^{i,j}_\kappa(\ell)}{\partial p^5_0}&= -\int_{-\infty}^0 dx \, \mathcal{K}(x,\ell) \left[ 2\frac{\Omega^{GR}_\Lambda(x)}{\Omega^{GR}_\Lambda(x=0)}\frac{\ell^2}{\chi_{GR}^2(x)}+3 \int_{-\infty}^x dx' \, I(x,x') \Omega^{GR}_M(x')\frac{\Omega^{GR}_\Lambda(x')}{\Omega^{GR}_\Lambda(x=0)}\frac{\ell^2}{\chi_{GR}^2(x')} \right] \nonumber \\
\frac{\partial f\sigma_8(x,k)}{\partial p^1_0}  &= -\int_{-\infty}^{x} G_f(x,x')\frac{\Omega^{GR}_\Lambda(x')}{\Omega^{GR}_\Lambda(x'=0)}dx' \nonumber \\
\frac{\partial f\sigma_8(x,k)}{\partial p^2_0}  &= -\int_{-\infty}^{x} G_f(x,x')\frac{\Omega^{GR}_\Lambda(x')}{\Omega^{GR}_\Lambda(x'=0)}k^2dx' \nonumber \\
\frac{\partial f\sigma_8(x,k)}{\partial p^3_0}  &= 2\int_{-\infty}^{x} G_f(x,x')\frac{\Omega^{GR}_\Lambda(x')}{\Omega^{GR}_\Lambda(x'=0)}k^2dx' \nonumber \\
\frac{\partial f\sigma_8(x,k) }{\partial p^4_0} &= -\int_{-\infty}^{x} G_f(x,x')\frac{\Omega^{GR}_\Lambda(x')}{\Omega^{GR}_\Lambda(x'=0)}dx' \nonumber \\
\frac{\partial f\sigma_8(x,k)}{\partial p^5_0}  &= -\int_{-\infty}^{x} G_f(x,x')\frac{\Omega^{GR}_\Lambda(x')}{\Omega^{GR}_\Lambda(x'=0)}k^2dx'.
\label{silvestriderivs}
\end{align}
\end{widetext}

%%%%%%%%%%%%%%%%%%%%%%%%%%%%%%%%%%%%%%%%%%%%%%%%%%%%%%%

% ---------------------- BIBLIOGRAPHY -----------------------------------------

%\bibliographystyle{h-physrev}
%\bibliography{pkappa_refs}

\end{document}